\documentclass[]{raa}
\usepackage{graphicx,times}
\usepackage{natbib}
\usepackage{amssymb,amsmath}
\bibpunct{(}{)}{;}{a}{}{,}
\usepackage[a4paper=true,dvipdfm=true,pagebackref=true]{hyperref}
\hypersetup{pdftitle = The title of my PDF, pdfauthor = My name, pdfsubject= The subject, pdfkeywords = keyword1 keyword2 keyword3}
\hypersetup{colorlinks = true, linkcolor = green, anchorcolor = red, citecolor = blue, filecolor = red, pagecolor = red, urlcolor = red}

\begin{document}

   \title{The YNEV stellar evolution and oscillation code$^*$
\footnotetext{\small $*$ Supported by the National Natural Science Foundation of China.}
}

 \volnopage{ {\bf 2014} Vol.\ {\bf X} No. {\bf XX}, 000--000}
   \setcounter{page}{1}

   \author{Q. S. Zhang\inst{1,2}}

   \institute{ Yunnan Observatories, Chinese Academy of Sciences,
              P.O. Box 110, Kunming 650011, China.; {\it zqs@ynao.ac.cn}\\
        \and
             Key Laboratory for the Structure and Evolution of Celestial Objects, Chinese Academy of Sciences, Kunming, 650011, China.\\
\vs \no
   {\small Received 2014 February 18; accepted 2014 June 26 }}

\abstract{We have developed a new stellar evolution and oscillation code YNEV, which calculates the structures and evolutions of stars, taking into account hydrogen and helium burning. A nonlocal turbulent convection theory and an updated overshoot mixing model are optional in this code. The YNEV code can evolve low- and intermediate-mass stars from pre-main sequence (PMS) to thermal pulsing asymptotic branch giant (TP-AGB) or white dwarf. The YNEV oscillation code calculates the eigenfrequencies and eigenfunctions of the adiabatic oscillations of given stellar structure. The input physics and the numerical scheme adopted in the code are introduced in this paper. The examples of solar models, stellar evolutionary tracks of low- and intermediate-mass stars with different convection theory (i.e., mixing-length theory (MLT) and the nonlocal turbulent convection theory), and stellar oscillations are shown.
\keywords{Stars --- Stellar structure --- Stellar evolution --- Stellar oscillations }}
   \authorrunning{Q.S. Zhang }            %author_head in even pages
   \titlerunning{The YNEV stellar evolution and oscillation code}  % title_head in odd pages
   \maketitle
%________________________________________________ sections below
%
\section{Introduction}           %% first-level sections will be auto-capitalized
\label{sect:intro}

We have developed a new stellar evolution and oscillation code named YNEV (YunNan EVolution code). This evolution code calculates the structures and evolutions of stars, taking into account hydrogen and helium burning. A feature of this code is that a nonlocal turbulent convection theory and an updated overshoot mixing model are optional to deal with the stellar turbulent convection. This evolution code can evolve a single low- and intermediate-mass star in spherical symmetry from PMS with the center temperature $T_C=10^5K$ to TP-AGB (for intermediate-mass stars) or white dwarf (for low-mass stars). The accessorial oscillation code calculates the eigenfrequencies and eigenfunctions of the adiabatic oscillations of given stellar structure. This paper introduces the input physics and the numerical scheme adopted in the code and shows the examples of solar models, stellar evolutionary tracks and stellar oscillations.

The contents of this paper are as follows: the physics involved in the YNEV code and their treatments are described in Section 2, the numerical calculations and the time / space step setting are introduced in Section 3, the generations of initial stellar models are introduced in Section 4, examples of stellar structures and evolutions and stellar oscillations are shown in Section 5, and possible improvements are discussed in Section 6.

\section{Input Physics}
\label{sect:inputphy}

\subsection{Equation of state}
\label{subsect:eos}

The equation of state (EOS) part in YNEV code calculates thermal functions pressure $P$, $(\partial ln P / \partial ln T )_{\rho}$, $(\partial ln P / \partial ln \rho )_{T}$, adiabatic temperature gradient $(\partial ln T / \partial ln P )_{S}$, and specific heat $c_P$ for input density $\rho$, temperature $T$ and chemical abundance $X_i$.

In low $T$ (in default $lgT<8.2$) and low $Z$ (in default $Z<0.045$) part, we use the EOS2005 tables \citep{RN02} to interpolate the thermal functions of gas, and then modify the thermal functions by taking into account the contributions of the radiative field which is assumed in local thermal equilibrium to the gas. In the calculations of gas thermal functions, we adopt the bicubic interpolation on $\rho$ and $T$ in order to obtain smooth (the first-order derivatives exist) thermal functions, and use linear interpolations on hydrogen abundance $X$ and $Z$ (for $Z$, quadratic interpolation between 3 tables with different metallicity is optional and this is usually used in solar models).

In high $T$ (in default $lgT>8.3$) or high $Z$ (in default $Z>0.05$) part, we assume a mixed ideal gas comprises ions, electrons, positrons and the radiative field. The thermal functions of the mixed ideal gas can be solved by using statistic physics (see, e.g., \citealp{ta99}). In order to reduce the time costs in the calculations of thermal functions of electrons and positrons, we interpolate them from a table calculated off-line. The table stores thermal functions of electrons and positrons in different rest electron number density $\rho_E$ (in $mol/cm^3$) and temperature $T$. In the calculations of the table, \citeauthor{apa98}'s \citeyearpar{apa98} scheme is adopted to calculate the integral of Fermi functions.

In the connecting region between the above parts, we interpolate thermal functions from two schemes above as follows. For a thermal function, say $A$, we denote the value of $A$ calculated using the first scheme to be $A_1$ and that using the second scheme to be $A_2$. We calculate the final result of $A$ by using the interpolation:
\begin{eqnarray}
&&A = a{A_1} + (1 - a){A_2}, \\ \nonumber
&&a = f(\lg T,8.2,8.3)f(Z,0.045,0.05), \\ \nonumber
&&f(x,{x_1},{x_2})  \equiv  \frac{1 }{2} \{{1 - \sin \{ \{ {\rm{max}}[0,{\rm{min}}(1,\frac{{x - {x_1}}}{{{x_2} - {x_1}}})] - \frac{1}{2}\} \pi \} }\}.
\end{eqnarray}%
It is obviously that the interpolation leads to a smooth thermal function $A$.

\subsection{Opacity}
\label{subsect:opac}

The opacity part of YNEV calculates the opacity for input density $\rho$, temperature $T$ and chemical abundance $X_i$. Similar to the EOS, the bicubic interpolation on $\rho$ and $T$ and linear interpolations on hydrogen abundance $X$ and $Z$ (for $Z$, quadratic interpolation is optional) are performed. The opacity tables are same to which are used in MESA \citep{pax11}. On the electron conduction opacity, \citet{cas07}, \citet{iben75} and \citet{yak80} are used. On the radiative opacity, the OPAL tables with fixed metal and CO enhanced \citep{ig96}, the OP tables \citep{OP05} and the F05 low-temperature tables \citep{F05} are used.

The choice of using tables with fixed metal or with CO enhanced is based on the input chemical abundance $X_i$.
Tables with fixed metal are adopted in the case of $Z\leq0.045$, tables with CO enhanced are adopted in the case of $Z\geq0.05$. A smooth interpolation similar to the EOS is used when $0.045<Z<0.05$. Since the YNEV code can be used for stars with initial $Z<0.04$ (higher $Z$ is not supported by the EOS tables, and we do not attempt to use the mixed ideal gas model in the envelope of a star), this scheme is in general reasonable. An inconsistence is in the start of helium burning, where CO enhanced tables should be used but fixed metal tables are actually used. This should not lead to unacceptable results, since the helium burning core is dominated by electron conduction opacity and the duration from the helium ignition to $Z=0.05$ is not long.

\subsection{Nuclear reaction}
\label{subsect:nucl}

In YNEV stellar evolution code, we trace the evolution of following elements:
$^1H$, $^2H$, $^3He$, $^4He$, $^7Li$, $^{12}C$, $^{13}C$, $^{14}N$, $^{15}N$, $^{16}O$, $^{17}O$.

The hydrogen and helium burnings (p-p chains, CNO cycles, $3\alpha$, $4\alpha$ and $5\alpha$ reactions) are taking into account:
\begin{eqnarray}
&&^1H{(p,v{e^ + })^2}H{(p,\gamma )^3}He, \\ \nonumber
&&^3He{(^3}He,pp){}^4He{;^3}He(\alpha ,\gamma ){}^7Be, \\ \nonumber
&&^7Be({e^ - },v){}^7Li(p,\alpha ){}^8Be(,\alpha ){}^4He, \\ \nonumber
&&^7Be(p,\gamma ){}^8B(v{e^ + }){}^8Be(,\alpha ){}^4He, \\ \nonumber
&&^{12}C(p,\gamma ){}^{13}N{(,v{e^ + })^{13}}C(p,\gamma ){}^{14}N(p,\gamma ){}^{15}O(,v{e^ + }){}^{15}N, \\ \nonumber
&&^{15}N(p,\gamma \alpha ){}^{12}C;{}^{15}N(p,\gamma ){}^{16}O(p,\gamma ){}^{17}F(,v{e^ + }){}^{17}O(p,\gamma \alpha ){}^{14}N, \\ \nonumber
&&^4He(\alpha \alpha ,\gamma ){}^{12}C(\alpha ,\gamma ){}^{16}O(\alpha ,\gamma ){}^{20}Ne, \\ \nonumber
&&^{14}N(\alpha ,\gamma ){}^{18}F,{}^{15}N(\alpha ,\gamma ){}^{19}F,{}^{13}C(\alpha ,n\gamma ){}^{16}O,{}^{17}O(\alpha ,n\gamma ){}^{20}Ne.
\end{eqnarray}%

The basic rates of all reactions above are based on \citet{ang99}, except for the $\beta$-decay of $^7Be$, which is based on \citet{cf88}.
The electron screening factors are based on \citet{sal54} (for weak screening) and \citet{dgc73} (for intermediate or strong screening) with a smooth interpolation.
The neutrinos energy-loss rates are calculated by using a public code by \citet{ito96}. In the calculation of chemical evolution due to nuclear reactions, we use implicit scheme to work out the nuclear reaction networks, and the abundance of $^7Be$ is assumed to be in equilibrium. In fully mixed burning zone (e.g., convective burning core), nuclear rates are integrated in the whole zone to calculate the variations of average chemical abundance. The initial composition in metal can be set to GN93 \citep{GN93}, GS98 \citep{GS98} or AGSS09 \citep{AGSS09}. The initial abundance of isotopes $D$ (i.e., $^2H$), $^3He$, $^7Li$, $^{13}C$, $^{15}N$, $^{17}O$ are based on \citet{AGSS09}.

\subsection{Convection}
\label{subsect:conv}

The convection in stellar interior plays an important role because the convection leads to entropy and chemical mixing which dominates the stellar structure and evolution. Unfortunately, the convection is still not very clear. At present, the local convection theory named Mixing-Length Theory (MLT) \citep{MLT58} is widely used in modeling stars. Although the MLT has some significant shortcomings (phenomenological theory, local theory being unable to study the convective overshoot, describing stellar turbulence in single length scale, etc.), it is very convenient to be implemented in code. More reasonable theories of stellar convection are the Turbulent Convection Models (TCMs) (e.g., \citealp{xio81,xio97,can97,can11,can98,den06,li07,li12}) which are based on fluid dynamics equations. The TCMs are of the capability to study the overshoot and show consistent results to helioseismology \citep{chr11}. On the other hand, in an overall view, the results of MLT theory are similar to the TCM except in the case of convective envelope and the convective stable region. In the stellar interior convection zones, both MLT and TCM show efficient entropy and chemical mixing so that temperature gradient is almost adiabatic and chemical elements are fully mixed.

In the standard version of YNEV code, the MLT is still adopted to deal with the convective entropy transport, the convective unstable zone (i.e., convection zone determined by Schwarzschild criterion) is artificially fully mixed, and the convective overshoot can be taking into account in the traditional way that the artificially fully mixed region is extended by $l_{OV}=\alpha_{OV}H_P$ where $H_P$ is pressure scale height and $\alpha_{OV}$ is a parameter. In the center helium burning phase, the induced semi-convection outside the convective core is implemented \citep{cas85}. The temperature gradient $\nabla$ determined by the standard MLT theory is calculated as follows (which may be not appeared before but is equivalent to other forms of standard MLT theory):
\begin{eqnarray}
 {f^3} + a{f^2} + (b - a)f - b = 0, \label{newMLT}
\end{eqnarray}%
where
\begin{eqnarray}
 f &=& \frac{{\nabla  - {\nabla _{ad}}}}{{{\nabla _R} - {\nabla _{ad}}}}, \\ \nonumber
 a &=& \frac{{23}}{4}{K^{ - 1}};b = (1 + \frac{{729}}{{64}}{K^{ - 1}}){K^{ - 1}}, \\ \nonumber
 K &=& \frac{1}{{32}}[{(\frac{{{\alpha}\rho {c_p}l}}{\lambda })^2}g\delta {H_P}({\nabla _R} - {\nabla _{ad}})], \label{newMLT2}
\end{eqnarray}%
and $l=\alpha H_P$ in which $l$ and $\alpha$ are mixing length and the MLT parameter.
A table of the roots $f$ within the accuracy of $10^{-13}$ with different values of $K$ in the range of $-5 \leq lgK  \leq 40$ was calculated off-line. When we need to solve the MLT, the table is read to get a good guess value of $f$ for input $K$, which leads to convergency in high accuracy after few Newton iterations. Outside the range of $-5 \leq lgK  \leq 40$, the following approximate solutions within the accuracy of $f$ being $10^{-12}$ is used:
\begin{eqnarray}
f &=& 1 - \frac{{64}}{{729}}{K^2}, lgK<-5; \\ \nonumber
f &=& {K^{ - \frac{1}{3}}}, lgK>40.
\end{eqnarray}%
It is not difficult to find those approximate solutions based on equation (\ref{newMLT}) in the behaviors of $K\rightarrow0$ and $K\rightarrow+\infty$.

In a beta version of YNEV code, the TCM developed by \citet{li07} can be used to replace the MLT to study the convective entropy transport, and the turbulent convective mixing model developed by \citet{zha13} can be used to study the convective overshoot mixing.

\subsection{Diffusion}
\label{subsect:dif}

The particle diffusion and gravitational settling, which are optional in the code, are calculated by solving Burger¡¯s equations with diffusion velocities / coefficients by \citet{tbl94}.
The elements are assumed to be fully ionized. The electrons are included. In default, $^1H$, $^3He$, $^4He$, $^{12}C$, $^{13}C$, $^{14}N$ and $^{16}O$ are taking into account, other elements are assumed to be $^{20}Ne$.

\subsection{Atmosphere}
\label{subsect:atm}

The atmosphere boundary conditions are based on the definition of effective temperature and the integral of the adopted $T-\tau$ relation $T=T(\tau)$.
In YNEV code, the outer boundary of stellar structure equations is set to be the location where $T=T_{eff}$. The definition of effective temperature gives a boundary condition $L_S=4 \pi r^2_S \sigma T^4_S$. Another boundary condition is based on the atmosphere (assumed to be homogeneous) integral:
\begin{eqnarray}
\frac{{d\ln \rho }}{{d\tau }} = \delta [\frac{g}{{P\kappa }}{(\frac{{\partial \ln T}}{{\partial \ln P}})_\rho } - \frac{{d\ln T}}{{d\tau }}],
\end{eqnarray}%
where $\delta=-(\partial ln \rho / \partial ln T )_P$ and $g=GM/R^2$, $g$ is assumed to be a constant in the atmosphere. The integral zone is from $\tau=0$ to $\tau=\tau_S$ where $\tau_S$ is defined by $(T_{eff}=)T_S=T(\tau_S)$, and we adopt the initial condition $\rho_{\tau=0}=10^{-10}$. Because the opacity tables are for $-8 \leq lgR \leq 8$ where $lgR=lg\rho-3lgT+18$, setting $\rho_{\tau=0}=10^{-10}$ is required for calculations of high temperature ($lgT>5$) white dwarfs. The second-order Runge-Kutta method is adopted in the numerical integral. The integral gives the value of density at the surface $\rho_S$. $ dlnT / d\tau $ and the value of $\tau_S$ are determined by the adopted $T-\tau$ relation. In YNEV code, there are two optional $T-\tau$ relations: the Eddington gray model and \citeauthor{ks66}'s \citeyearpar{ks66} $T-\tau$ relation.

\subsection{Mass loss}
\label{subsect:massloss}

In default, the mass loss is implemented in YNEV code by ejecting the stellar outer envelope with the mass $M_{env}=-dM/dt \cdot \Delta t$ where $dM/dt$ is the mass-loss rate and $\Delta t$ is the time step between the current stellar model and the previous one. Another optional scheme is to evaporate the envelope that modifying the independent variable ($ln[M_r/(M-M_r)]$) on all mesh points to result a mass-loss of $10\%$ (in default) between two mesh points in the envelope. The evaporation scheme is better for convergency. The ejection scheme is more reasonable in physics because the evaporation scheme leads to errors on the gravitational energy release (i.e., the $TdS/dt$ term in stellar structure equations), but it requires smaller time step for convergency. If we do not care the details of the gravitational energy release in the envelope caused by mass-loss, the evaporation scheme is also valid and reduces the calculation time cost. There are three options of the mass-loss rate: \citet{rei75}, \citet{wal84}, \citet{dej98}. It is not difficult to implement other expression of mass-loss rate.

\section{Numerical calculation}
\label{sect:num}

\subsection{Numerical Scheme}
\label{subsect:numsch}

The code assumes that the star is one-dimensional and in hydrostatic, and ignores the effects of rotation. The evolution of the element abundances in stellar interior and the stellar structure equations are solved alternatively. The evolution of the element abundances is calculated based on the previous stellar structure, then the new structure is determined by the updated element abundances profile. Although this scheme may lead to self-consistent problems, the errors are small in the most cases since the time step is not large enough.

The stellar structure equations are written on the form as follows:
\begin{eqnarray}
\frac{{d\lg P}}{{dq}} + \frac{{Mm(1 - m)}}{{\ln 10}}\frac{g}{{4\pi {r^2}P}} = 0, \label{stellar1}
\end{eqnarray}%
\begin{eqnarray}
\frac{{d\lg T}}{{dq}} + \frac{{Mm(1 - m)}}{{\ln 10}}\frac{{g\nabla }}{{4\pi {r^2}P}} = 0,
\end{eqnarray}%
\begin{eqnarray}
\frac{{d\lg r}}{{dq}} - \frac{{Mm(1 - m)}}{{\ln 10}}\frac{1}{{4\pi {r^3}\rho }} = 0,
\end{eqnarray}%
\begin{eqnarray}
\frac{{d{l_r}}}{{dq}} - \frac{{Mm(1 - m)}}{{{L_0}}}[{\varepsilon _N} - {\varepsilon _\nu } - ({c_P}\frac{{\partial T}}{{\partial t}} - \frac{\delta }{\rho }\frac{{\partial P}}{{\partial t}})] = 0, \label{stellar4}
\end{eqnarray}%
where the independent variable $q=ln[m/(m-1)]$, $m=M_r/M$ is the mass fraction, $l_r=L_r/L_0$ is dimensionless luminosity, and $L_0=Max(L_r)$ is the maximum value of luminosity in stellar interior. $\nabla$ is the temperature gradient determined by convection theory, e.g., for the MLT theory, $\nabla$ is calculated by using Equations (\ref{newMLT}) and (\ref{newMLT2}). The boundary conditions are as follows:
for the inner boundary where $r=r_1$:
\begin{eqnarray}
{r_1}^3 = \frac{{3{m_1}}}{{4\pi {\rho _1}}},
\end{eqnarray}%
and
\begin{eqnarray}
{l_{r,1}} = \frac{{{m_1}}}{{{L_0}}}{[{\varepsilon _N} - {\varepsilon _\nu } - ({c_P}\frac{{\partial T}}{{\partial t}} - \frac{\delta }{\rho }\frac{{\partial P}}{{\partial t}})]_1},
\end{eqnarray}%
and for the outer boundary where $r=r_N$:
\begin{eqnarray}
{l_{r,N}}{L_0} = 4\pi {r_N}^2\sigma {T_N}^4; (T_N=T_{eff}),
\end{eqnarray}%
and
\begin{eqnarray}
{\rho _N}={\rho _S} \label{stellar2},
\end{eqnarray}%
where $\rho _S$ is determined by the atmosphere integral described in Section \ref{subsect:atm}.

The chemical evolution equations in the stellar interior are in general the diffusion equation:
\begin{eqnarray}
\frac{{\partial X}}{{\partial t}} + b\frac{{\partial (aF)}}{{\partial q}} = b\frac{\partial }{{\partial q}}({a^2}bD\frac{{\partial X}}{{\partial q}}) + R;\{ a = \frac{{dm}}{{dr}},b = \frac{{dq}}{{dm}}\},
\end{eqnarray}%
where $X$ is the chemical abundance vector, $F$ is the flux vector, $D$ is the diffusion coefficients matrix, and $R$ is the nuclear reaction rates vector. When the settling is taken into account, $F$ and $D$ are calculated based on \cite{tbl94}. The convective / overshoot mixing can be represented by adding a diffusion coefficient on the diagonal components in $D$. This equation can be rewrite as two first-order equations by defining $W$ as the total diffusion flux vector:
\begin{eqnarray}
b\frac{{\partial (aW)}}{{\partial q}} - (R - \frac{{\partial X}}{{\partial t}}) = O, \label{diff1}
\end{eqnarray}%
\begin{eqnarray}
abD\frac{{\partial X}}{{\partial q}} - (F - W) = O. \label{diff1a}
\end{eqnarray}%
The boundary conditions for the diffusion equations are:
\begin{eqnarray}
W = O \label{diff2},
\end{eqnarray}%
at the center and the stellar surface.

The radial part of the stellar adiabatic oscillation equation is a linear equation:
\begin{eqnarray}
\frac{{\partial J}}{{\partial \ln r}} - A(r,\omega)J = O, \label{osc1}
\end{eqnarray}%
where $J=(\xi_r,P',\Phi',g')^T$ is the vector determining the properties of the stellar oscillations, $A$ is the coefficient matrix of the oscillation equation, and $\omega$ is the frequency. The elements of matrix $A$ can be found in oscillation literature. We have adopted the dimensionless form, see, e.g., \citet{li10}. Two boundary conditions are at the center and other two boundary conditions are at the surface. Four boundary conditions are all linear and homogeneous.

The stellar structure equations (\ref{stellar1}-\ref{stellar2}), the diffusion equations (\ref{diff1}-\ref{diff2}) and the stellar adiabatic oscillation equations (\ref{osc1}) are all first-order equations with two points boundary conditions. We use Newton iterations (linearization) method to solve the two points boundary conditions problems (\ref{stellar1}-\ref{stellar2}) and (\ref{diff1}-\ref{diff2}). The implicit discretisation is adopted for the time derivative in equations (\ref{stellar4}, \ref{diff1} \& \ref{diff1a}). The equation (\ref{osc1}) and its boundary conditions are already linear. The general form of those problem is:
\begin{eqnarray}
H(\frac{{\partial U}}{{\partial q}},U,q) = O,
\end{eqnarray}%
with the boundary conditions at $q_1$ and $q_N$:
\begin{eqnarray}
B(U_1,q_1) = O,
\end{eqnarray}%
\begin{eqnarray}
C(U_N,q_N) = O,
\end{eqnarray}%
where $q$ is the independent variable, $U$ is the dependent variables vector with $n$ elements, $H$ is the vector of zero functions, $B$ is the vector of zero functions with $n_1$ elements, $C$ is the vector of zero functions with $n_2$ elements, and $n=n_1+n_2$. The two point 2nd order discretisation is adopted. For the diffusion equation, we adopted the conservation form, in which the flux and the chemical abundance are not in the same mesh points, to ensure a correct flux, i.e., we set $W_k$ to be $W$ at the middle point between mesh point $k$ and $k+1$. In general, the first-order equation between mesh point $k$ and $k+1$ is as follow:
\begin{eqnarray}
{H_k} = {H_k}({U_k},{U_{k + 1}}) = O; k=1,2,...,N-2,N-1.
\end{eqnarray}%
We use Newton iterations method to solve these equations on all mesh points. Expanding ${H_k}({U_k},{U_{k + 1}})$ on the guess solution ${U^{(i)}}_k$ (for the stellar structure equations and the diffusion equations, which are time-dependent, the guess solutions ${U^{(0)}}$ are set as the values in previous model, and for the stellar oscillations, the guess solution is ${U^{(0)}}=O$) with ignoring high-order term, we found:
\begin{eqnarray}
\frac{{\partial {H_k}}}{{\partial {U_k}}}\Delta {U_k} + \frac{{\partial {H_k}}}{{\partial {U_{k + 1}}}}\Delta {U_{k + 1}} =  - {H_k}({U^{(i)}}_k,{U^{(i)}}_{k + 1}); k=1,2,...,N-2,N-1, \label{numeq1}
\end{eqnarray}%
where ${\partial {H_k}/\partial {U_k}}$ and ${\partial {H_k}/\partial {U_{k+1}}}$ are Jacobi matrices, and a revised solution:
\begin{eqnarray}
{U^{(i + 1)}}_k = {U^{(i)}}_k + \Delta {U_k},
\end{eqnarray}%
For the boundary conditions, the similar results are:
\begin{eqnarray}
\frac{{\partial B}}{{\partial {U_1}}}\Delta {U_1} =  - B({U^{(i)}}_1), \label{numeq2}
\end{eqnarray}%
\begin{eqnarray}
\frac{{\partial C}}{{\partial {U_N}}}\Delta {U_N} =  - C({U^{(i)}}_N). \label{numeq3}
\end{eqnarray}%
Equations (\ref{numeq1}), (\ref{numeq2}) and (\ref{numeq3}) are the complete equations for calculating the corrections $\Delta {U_k}$. Those equations are equivalent to a linear equation with a huge coefficient matrix with only the elements at/near the diagonal being nonzero. It is not difficult to solve the equation by using the method of forward eliminations and backward recursions. When the corrections are not small enough, the zero function $H$ are therefore not close enough to zero. We then repeat this process until the corrections are in allowed errors. Typically, we set the accuracies as $Max(\mid \delta lg\rho\mid ,\mid \delta lgT\mid ,\mid \delta lgr\mid ,\mid \delta l_r\mid )<10^{-6}$ in the stellar structure equations, and $\mid \delta X(^1H) \mid<10^{-8}$, $\mid \delta X(^4He) \mid<10^{-8}$ and $\mid \delta X_i /Max(10^{-10},X_i) \mid<10^{-8}$ for other chemical elements in the chemical evolution equation. The oscillation equation is already linear, and it is equivalent to the equations (\ref{numeq1}-\ref{numeq3}) with ${U^{(0)}}=O$. We only need to solve the linear equations once.

In scanning the eigenfrequencies in the oscillation equation, we define the discriminant $V(\omega)$ as the determinant of the coefficient matrix of the equation on an arbitrary mesh point $k_0$ (in default $k_0=N$):
\begin{eqnarray}
P_{k_0}\Delta {U_{k_0}}=Q_{k_0}; V(\omega)=Det(P_{k_0}),
\end{eqnarray}%
where the coefficient matrix $P_{k_0}$ and the vector $Q_{k_0}$ can be worked out in the process of forward eliminations and backward recursions. $\omega$ is an eigenfrequency when $V(\omega)=0$, since the homogeneous equation have nonzero solutions only if the coefficient matrix has singularity. This definition ensures the continuity of the discriminant $V(\omega)$ and makes conveniences for scanning the eigenfrequencies. In solving the eigenfunctions $U$ for a validated eigenfrequency $\omega_i$, we use an inhomogeneous boundary condition (e.g., $\xi_r=1$) to replace an homogeneous boundary condition in order to eliminate the singularity of the coefficient matrix.

\subsection{Numerical scheme: implement of the Turbulent Convection Model}
\label{subsect:numtcm}

In a beta version of the YNEV evolution code, the Turbulent Convection Model (TCM) developed by \citet{li07} can be replace the MLT theory and used to calculate the stellar turbulent variables. The TCM equations are as follows:
\begin{eqnarray}
\frac{2}{{\rho {r^2}}}\frac{\partial }{{\partial r}}(\rho {r^2}{C_s}{k_r}\tau \frac{{\partial {k_r}}}{{\partial r}}) = \frac{1}{3}k{\tau ^{ - 1}} - \frac{{\delta g}}{T}\overline {{u_r}'T'}  + {C_k}{\tau ^{ - 1}}({k_r} - \frac{k}{3}), \label{tcm1}
\end{eqnarray}%
\begin{eqnarray}
\frac{2}{{\rho {r^2}}}\frac{\partial }{{\partial r}}(\rho {r^2}{C_s}{k_r}\tau \frac{{\partial k}}{{\partial r}}) = k{\tau ^{ - 1}} - \frac{{\delta g}}{T}\overline {{u_r}'T'}, \label{tcm2}
\end{eqnarray}%
\begin{eqnarray}
\frac{4}{{\rho {r^2}}}\frac{\partial }{{\partial r}}(\rho {r^2}{C_{t1}}{k_r}\tau \frac{{\partial \overline {{u_r}'T'} }}{{\partial r}}) =  - \frac{{\delta g}}{T}\overline {T'T'}  - 2{k_r}\frac{T}{{{H_P}}}(\nabla  - {\nabla _{ad}}) + {C_t}(1 + {P_e}^{ - 1}){\tau ^{ - 1}}\overline {{u_r}'T'}, \label{tcm3}
\end{eqnarray}%
\begin{eqnarray}
\frac{1}{{\rho {r^2}}}\frac{\partial }{{\partial r}}(\rho {r^2}{C_{e1}}{k_r}\tau \frac{{\partial \overline {T'T'} }}{{\partial r}}) =  - \overline {{u_r}'T'} \frac{T}{{{H_P}}}(\nabla  - {\nabla _{ad}}) + {C_e}(1 + {P_e}^{ - 1}){\tau ^{ - 1}}\overline {T'T'}, \label{tcm4}
\end{eqnarray}%
where $k_r=\overline{u_r'u_r'}/2$ is the radial turbulent kinetic energy, $k$ is the turbulent kinetic energy,
$\overline{u_r'T'}$ describes the convective heat flux, $\overline{T'T'}$ is the turbulent temperature variance, $\tau = k/\varepsilon$ is the dissipation timescale with the turbulent dissipation rate $\varepsilon=k^{3/2}/l$ and $l=\alpha_{TCM} H_P$, and $P_e=lk^{1/2}/D_R$ is the P\'{e}clet number with radiative diffusion coefficient $D_R=\lambda/(\rho c_P)$. $C_s$, $C_{t1}$ and $C_{e1}$ are dimensionless diffusion coefficients, $\alpha_{TCM}$, $C_{t}$ and $C_{e}$ are dimensionless dissipation coefficients, and $C_{k}$ is a parameter dominates the rate of $k_r/k$.

The default values for the parameters in the TCM are as follows \citep{zhl12,zha12}: $C_s=0.08$, $C_t=7.5$, $C_e=0.2$, $C_k=2.5$, $C_{t1}=0.02$ or $0$, $C_{e1}=0.02$ or $0$. The turbulent kinetic dissipation parameter $\alpha_{TCM}=0.8$ is based on the solar calibration with Eddington gray atmosphere model, or $\alpha_{TCM}=1.0$ for K-S atmosphere model. Solar calibrations for different compositions shows $\alpha=(2.1 \sim 2.2)\alpha_{TCM}$.

We solve the stellar structure equations and the TCM equations alternately to find the solution satisfying both equations. The TCM equations are solve by using an iteration method based on multigrid method.
In numerical solving the TCM equations, the variables (except the temperature gradient $\nabla$) are based on current stellar structure.
The following equation is substituted into the TCM equations for the temperature gradient $\nabla$:
\begin{eqnarray}
\nabla  = {\nabla _{R,therm}} - \frac{{{H_P}}}{T}\frac{{\rho {c_P}\overline {{u_r}'T'} }}{\lambda }, \label{grt}
\end{eqnarray}%
where $\nabla _{R,therm}$ is the radiative temperature gradient for thermal energy flux \citep{zha14}:
\begin{eqnarray}
{\nabla _{R,therm}} = {\nabla _R} - \frac{{{H_P}}}{T}\frac{{{F_K}}}{\lambda },
\end{eqnarray}%
and ${F_K}$ is the turbuleng kinetic energy flux calculated as follows:
\begin{eqnarray}
{F_K} =  - 2{C_s}\rho{k_r}\tau \frac{{\partial k}}{{\partial r}}.
\end{eqnarray}%
In solving the TCM equations, $\nabla _{R,therm}$ is determined by current stellar structure and previous turbuleng kinetic energy flux.

The steps of the implement of TCM in the code have been described by \cite{zha12}:

1. Solve the TCM equations based on the current stellar structure and previous turbuleng kinetic energy flux. Calculate the temperature
gradient $\nabla$ at all mesh points according to Equation (\ref{grt}).

2. Solve the localized TCM in which the diffusion terms are ignored. Calculate the corresponding
temperature gradient $\nabla_L$ at all mesh points.

3. Calculate the ratio $\eta=\nabla/\nabla_L$ at all mesh points. And calculate a relaxed $\eta'=\eta'_{pre}+\xi (\eta-\eta'_{pre})$, where the subscript 'pre' means previous values. The relaxation parameter $\xi$ is 0.618 in default.

4. Solve the stellar structure equations in which the temperature
gradient is calculated as $\nabla=\eta' \nabla_L$, and update the stellar
structure. $\nabla_L$ is calculated by the localized TCM in solving the stellar
structure equations.

5. Check the differences $\mid \eta- \eta' \mid$ and $\mid \eta- \eta_{pre} \mid$. The calculations
are thought to converge if both differences are less
than the required accuracy ($10^{-3}$ in default) at all mesh points; otherwise,
return to step 1.

Although the relaxation improves the numerical stability and ensures that the scheme works in most cases of stellar evolutions, it should be mentioned that this implement of TCM still does not work in some cases. There is still some numerical problems for the implement of TCM.

When the TCM is adopted, the updated overshoot mixing model by \citet{zha13} is also adopted in solving the stellar chemical evolution.
The diffusion coefficient for mixing in overshoot region in this model is as follows:
\begin{eqnarray}
D_{OV}=C_{OV}\frac{\varepsilon}{N_{turb}^2},
\end{eqnarray}%
where $N_{turb}^2$ is calculated as follows:
\begin{eqnarray}
{N_{turb}}^2 =  - \frac{{\delta g}}{{{H_P}}}[\nabla  - {\nabla _{ad}} - {C_1}{C_A}\sum\limits_{k = 1}^M {{{(\frac{{\partial \ln T}}{{\partial {X_k}}})}_{P,\rho ,X - \{ {X_k}\} }}\frac{{d{X_k}}}{{d\ln P}}} ],
\end{eqnarray}%
where $C_A=C_e+C_{OV}$ and, according to \citet{can11}, $C_1=\sigma_t=0.72$. The turbulent dissipation rate $\varepsilon$ is calculated by using the TCM. It is optional to use the exact representation of the diffusion coefficient \citep[see][Equation (26)]{zha13}, but there is no obvious difference. The only parameter, i.e., the dimensionless diffusion coefficient $C_{OV}$ is suggested as $C_{OV}\sim 10^{-3}$ based on the tests of solar model and the restriction of classical overshoot length being less than $0.4H_P$ \citep{zha13}, and the calibrations on effective temperatures and radii for low-mass eclipsing binary stars \citep{mz14}.

\subsection{Time step}
\label{subsect:timestep}

In the normal case, the time step in the calculations of stellar evolution depends on the following factors: the maximum correction in the Newton iterations, the variations of $^1H$ and $^4He$ abundance in the center, the ratio of burned $^1H$ in the previous time step to $^1H$ abundance in the center, the ratio of burned $^1H$ (and $^4He$) in the previous time step to $^1H$ (and $^4He$) abundance in burning shells, the ratio of burned minor elements ($D$, $^3He$, $^7Li$ and $^{12}C$) in the previous time step to their abundances in the center in PMS stage, and the variations of $lgT_{eff}$ and $lgL$ between previous two stellar models.

When a stellar structure is converged in the Newton iterations, the time step for the next stellar model is estimated by taking into account those factors. In the calculation of the next stellar structure, if there is no convergency, the time step is reduced and re-do the calculations. If there is still no convergency after too many (50, in default) times reducing time step and re-calculations, the code checks whether helium flashes in the stellar model. In the helium flash case, we don't attempt to trace the changes of the stellar structure and let the star jump to ZAHB model. In the case of no helium flash, the code stops.

\subsection{Space step}
\label{subsect:spacestep}

The mesh points in the calculation of stellar structure are controlled by the the differences of $lg\rho$, $lgT$, $r/R$, $L_r/L_{Max}$, $lg\tau$, $ln[M_r(/M-M_r)]$ (the independent variable), $X_i$ (chemical abundance of all elements) between two adjacent grids. Near the Schwarzschild convective boundaries and the boundaries of artificially fully mixed regions, the density of mesh points are extra added to be $5-10$ times of the normal case. This is designed to ensure the accuracy of the location of convective boundaries and the boundaries of fully mixed regions, which may sensitively affect the stellar evolution.

In the oscillation code, the mesh points of the input stellar model are all taken into account. However, especially for the high node number oscillation modes, the mesh points in the stellar model may be not dense enough in short wavelength regions, thus the difference of radius between two adjacent grids is much longer than the wavelength. This leads to the problem that the resolution is not enough to reveal the wave. Therefore, in the calculation of each oscillation mode, extra mesh points are temporarily added by linear interpolation to ensure that there are at leats five mesh points in a wavelength, where the wavelength is estimated by using the dispersion relation.

\subsection{Time costs}
\label{subsect:timecost}

The time cost for the YNEV code calculating stellar evolutionary models depends on the time / space step settings, stellar parameters (i.e., stellar mass), software (adopted fortran compiler) and the hardware. For a computer with two cores $\sim3GHz$ CPU and compiled by using the Intel Fortran compiler, with the number of mesh points $1000\sim1500$ and the number of time grids $\sim1100$, YNEV evolving a intermediate-mass star from the PMS to the helium burning out in the core costs about five minutes. With the same time / space step settings, YNEV evolving a low-mass star from the PMS to a low temperature white dwarf costs more than three hours, and the total number of time grids is about $\sim30000$. The most time costing phase is from the TP-AGB.

The most time costing part in the YNEV code is to solve the nuclear reaction networks on all mesh point. The opacity and EOS are interpolated from the tables with different metallicites because $Z$ slightly changes in the hydrogen burning region. If we ignore this slight change and calculate opacity and EOS based on the tables with fixed $Z$, the time costs should be reduced. We have tested the time costs by using the nuclear reaction part of \citeauthor{pacz69}'s \citeyearpar{pacz69} code to replace the nuclear reaction networks of YNEV and calculating opacity and EOS based on fixed Z. It is found that the time costs are significantly reduced, e.g., the case of the mediate-mass star mentioned above costs only a half minute. In order to ensure the accuracy, we usually prefer to use the nuclear reaction networks and interpolate opacity and EOS from tables with different metallicites.

To use the nonlocal TCM significantly increases the time costs. The reason is that, for all stellar models in the evolutionary series, the stellar structure equations and the TCM equations are solved alternately to find the final solution satisfying both equations. The case of the intermediate-mass star mentioned above costs near an hour when we adopt the nonlocal TCM to deal with the stellar convection.

\section{Initial model}
\label{subsect:inimod}

The YNEV code evolves star from PMS with center temperature $T_C=10^5K$, ZAMS or ZAHB. A database of initial PMS and ZAHB models with different stellar masses (both helium core mass and hydrogen envelope mass, for ZAHB models), metal compositions and fixed MLT parameter $\alpha$ in the cases of X=0.7, Z=0.02 and X=0.75, Z=0 are calculated off-line and stored.

The initial PMS model for given input stellar parameters is obtained by reading a stored PMS model with the same metal composition and the closest stellar mass and using Newton iterations method to solve the stellar structure with required stellar parameters. Relaxations are automatically performed if there is no convergency in Newton iterations, i.e., the stellar parameters are gradually changed from the stored PMS model to required values.
The initial ZAHB model for given input stellar parameters is obtained by using a similar way.
For the ZAMS model, in order to ensure the accuracy of the compositions in metal in stellar interior (since they may change via nuclear burning in the PMS stage), we do not used off-line calculated models. The adopted method to generate ZAMS model is to quickly evolve star from PMS to ZAMS (defined by $X_S-X_C=0.001$) by using half number of mesh points and twice time steps.

\section{Samples of stellar evolution and oscillations}
\label{sect:sample}

\subsection{Solar models}
\label{subsect:solar}

\begin{figure*}
\centering
\includegraphics[width=11cm]{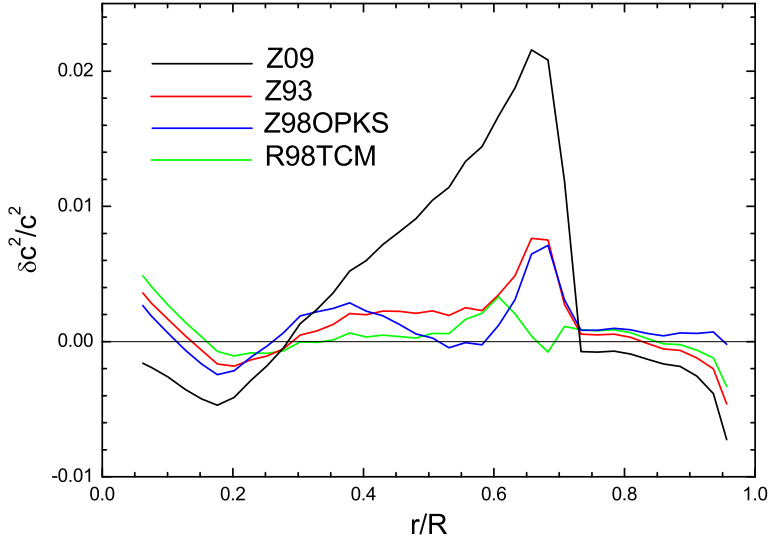}\vspace{1mm}
\caption{Differences of sound speed between models and helioseismic inversions. } \label{solar}
\end{figure*}

We use YNEV code to calculate four solar models: Z93, Z09, Z98OPKS and R98TCM. The models are evolved from ZAMS to the solar age $4.57Gyr$. Their radii and the luminosity are calibrated to $R_{\odot}=6.96\times10^{10}cm$ and $L_{\odot}=3.846\times10^{33}erg/s$ in an accuracy of $10^{-4}$. In model Z93, OPAL opacity and Eddington gray model are used, the solar composition GN93 is adopted and the ratio of metallicity to hydrogen $(Z/X)_S$ is calibrated to be 0.0245 \citep{GN93}. In model Z09, OPAL opacity and Eddington gray model are used, the solar composition AGSS09 is adopted and $(Z/X)_S$ is calibrated to be 0.0181 \citep{AGSS09}. In model Z98OPKS, OP opacity and K-S atmosphere model are used, the solar composition GS98 is adopted and $(Z/X)_S$ is calibrated to be 0.023 \citep{GS98}. The MLT is applied in Z93, Z09 and Z98OPKS models. In model R98TCM, GS98 composition \citep{GS98}, OPAL opacity and Eddington gray model are used, the TCM and the updated convective overshoot mixing model are adopted, the base of the convective envelope is calibrated to be $R_{bc}/R_{\odot}=0.7135$. The key information of the solar models are listed in Table 1. The comparison of sound speed between models and helioseismic inversions \citep{ba09} are shown in Fig.\ref{solar}. Comparing with the MESA \citep{pax11}, the YNEV solar model Z98OPKS shows almost the same results on the parameters, e.g., for MESA, $X_0=0.7065$, $Z_0=0.0191$, $Y_s=0.2433$, $Z_s=0.0170$, $R_{bc}/R_{\odot}=0.7140$, and the differences of sound speed for MESA \citep[see][Figure 21]{pax11} are also similar with the YNEV.

\begin{table*}
\centering
\caption{Parameters of the solar models. }
\begin{tabular}{lccccc}
\hline\hline\noalign{\smallskip}
Model  & Z93 & Z09 & Z98OPKS & R98TCM \\
\hline\noalign{\smallskip}
Atmosphere   & EG     & EG     & KS     & EG      \\
Composition  & GN93   & AGSS09 & GS98   & GS98    \\
Opacity      & OPAL   & OPAL   & OP     & OPAL    \\
$\alpha$     & 1.744  & 1.645  & 2.277  & 0.8069 (TCM)\\
$X_0$   & 0.7050 & 0.7179 & 0.7079 & 0.6998 \\
$Z_0$   & 0.0201 & 0.0152 & 0.0189 & 0.0200 \\
$Y_s$   & 0.2449 & 0.2359 & 0.2434 & 0.2565 \\
$Z_s$   & 0.0181 & 0.0136 & 0.0170 & 0.0186 \\
$(Z/X)_s$  & 0.0245 & 0.0181 & 0.0230 & 0.0257 \\
$R_{bc}/R_{\odot}$  & 0.7136 & 0.7248 & 0.7140 & 0.7135 \\
$[Li]/[Li]_0$  & 5.7\% & 14.8\% & 8.2\% & 1.7\% \\
\hline
\end{tabular}
\end{table*}

\subsection{Evolutionary tracks in the the HR Diagram: for the classical MLT theory}
\label{subsect:trackmlt}

\begin{figure*}
\centering
\includegraphics[width=11cm]{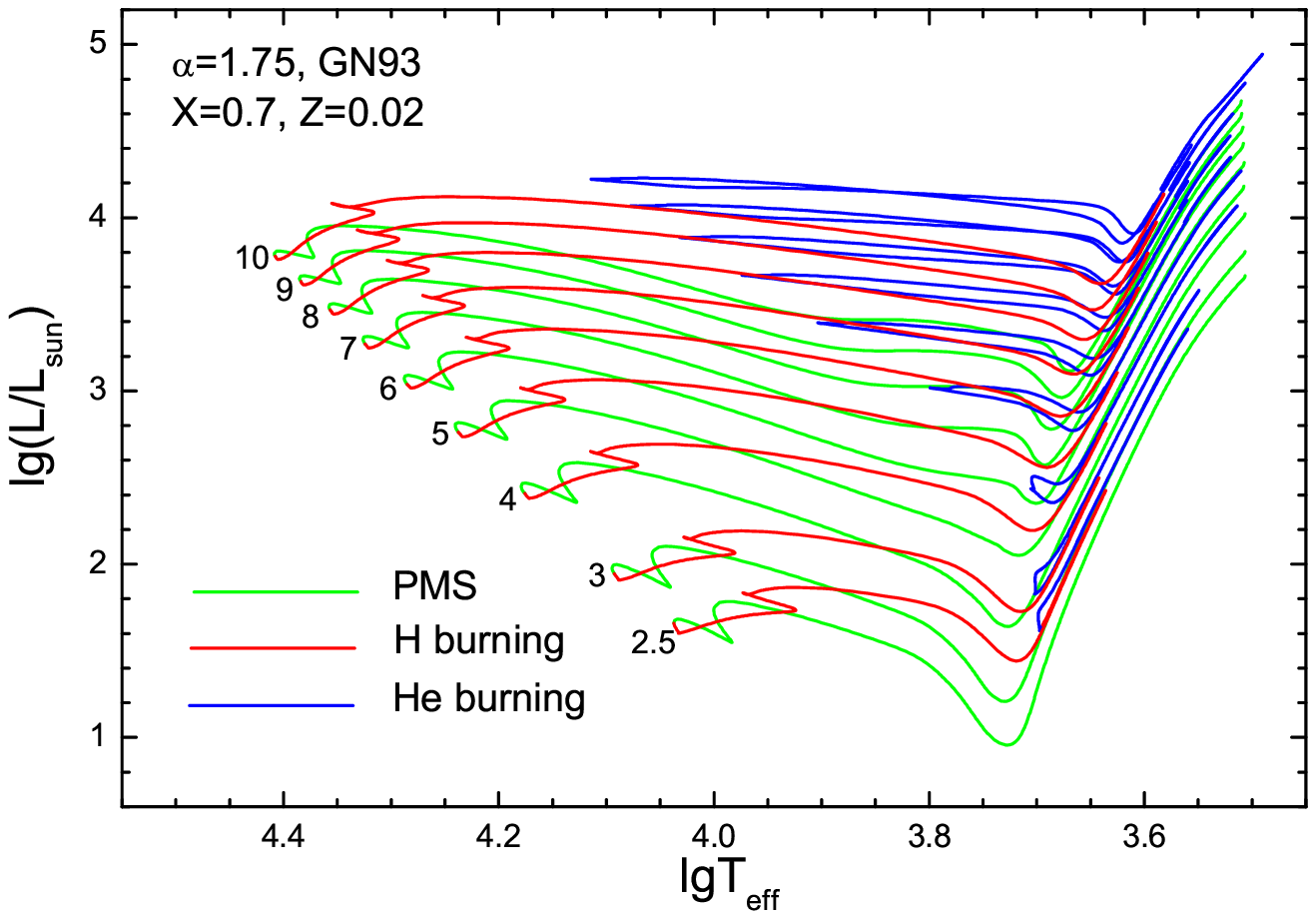}\vspace{1mm}
\caption{Evolutionary tracks of intermediate-mass stars with X=0.7, Z=0.02 and MLT parameter $\alpha=1.75$. Metal composition is same to GN93. In the tracks, the green, red and blue parts correspond to PMS, hydrogen burning and helium burning phases. The numbers near the ZAMSs indicate the stellar mass (in solar mass). } \label{X7Z2d-2}
\end{figure*}

\begin{figure*}
\centering
\includegraphics[width=11cm]{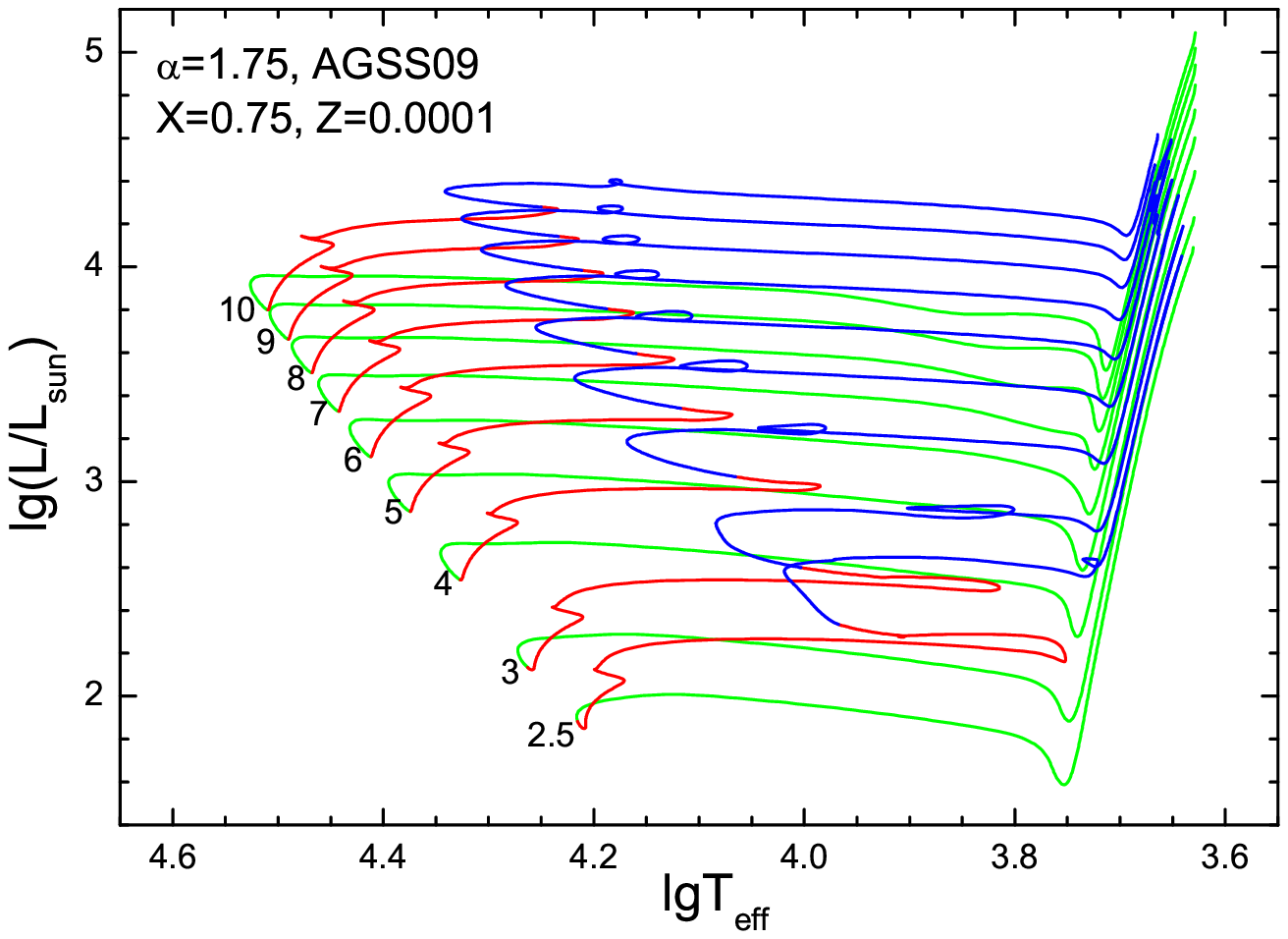}\vspace{1mm}
\caption{Similar to Fig.2. But for X=0.75, Z=0.0001 and AGSS09 metal composition. } \label{X75Z1d-4}
\end{figure*}

Evolutionary tracks of intermediate-mass stars with mass between $2.5 \leq M/M_{\odot} \leq 10$ generated by YNEV code are shown in Fig.\ref{X7Z2d-2} and Fig.\ref{X75Z1d-4}. Two Figures are of different stellar chemical composition. Stars evolve from the PMS with $T_C=10^5K$ to AGB stage. The PMS stage, the hydrogen burning stage (defined by $X_S-X_C\geq0.001$ and $X_C+Y_C\geq0.95$) and the helium burning stage (defined by $X_C+Y_C\leq0.95$) are shown as green, red and blue lines, respectively. The YNEV evolutionary tracks are similar with the tracks calculated by the FRANEC code. The blue loops for the intermediate-mass stars are sensitive results in stellar evolutionary codes. Comparing Figure \ref{X7Z2d-2} with the FRANEC evolutionary tracks \citep[see][Figure 3]{FRANECBlueloop}, it can be found that the blue tips of the blue loops in YNEV and FRANEC are almost at the same locations: $(lgT\approx4.12,lgL\approx4.2)$ for $10M_{\odot}$ star, $(lgT\approx4.03,lgL\approx3.9)$ for $8M_{\odot}$ star, $(lgT\approx3.97,lgL\approx3.65)$ for $7M_{\odot}$ star, $(lgT\approx3.90,lgL\approx3.4)$ for $6M_{\odot}$ star, and $(lgT\approx3.78,lgL\approx3.0)$ for $5M_{\odot}$ star.

\begin{figure*}
\centering
\includegraphics[width=11cm]{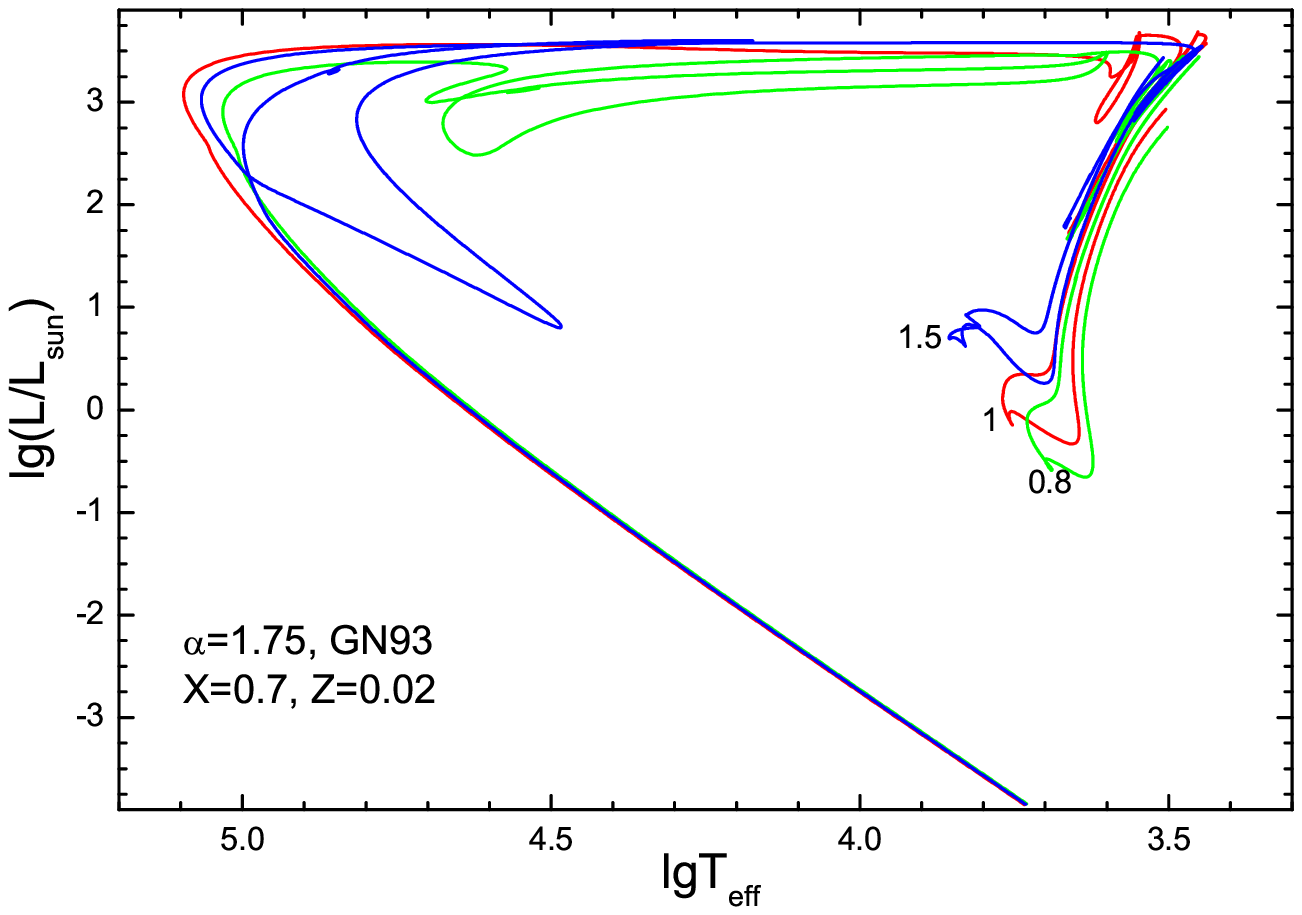}\vspace{1mm}
\caption{Evolutionary tracks of $1.5M_{\odot}$, $1M_{\odot}$ and $0.8M_{\odot}$ stars with X=0.7, Z=0.02 and MLT parameter $\alpha=1.75$. Metal composition is same to GN93. } \label{low-mass}
\end{figure*}

Three samples of the evolutionary tracks of low-mass stars from PMS to cooling series white dwarf models are shown in Fig.\ref{low-mass}. When the helium flashes, the locations of the stars automatically jump to ZAHB without tracing the helium flash process in detail. After the end of center helium burning, \citeauthor{rei75}'s \citeyearpar{rei75} mass-loss rate are adopted. For the $1.5M_{\odot}$ star, the mass-loss rate is enhanced by a factor of 5. The stars loss their envelope in AGB stage and finally evolve to white dwarfs.

\begin{figure*}
\centering
\includegraphics[width=11cm]{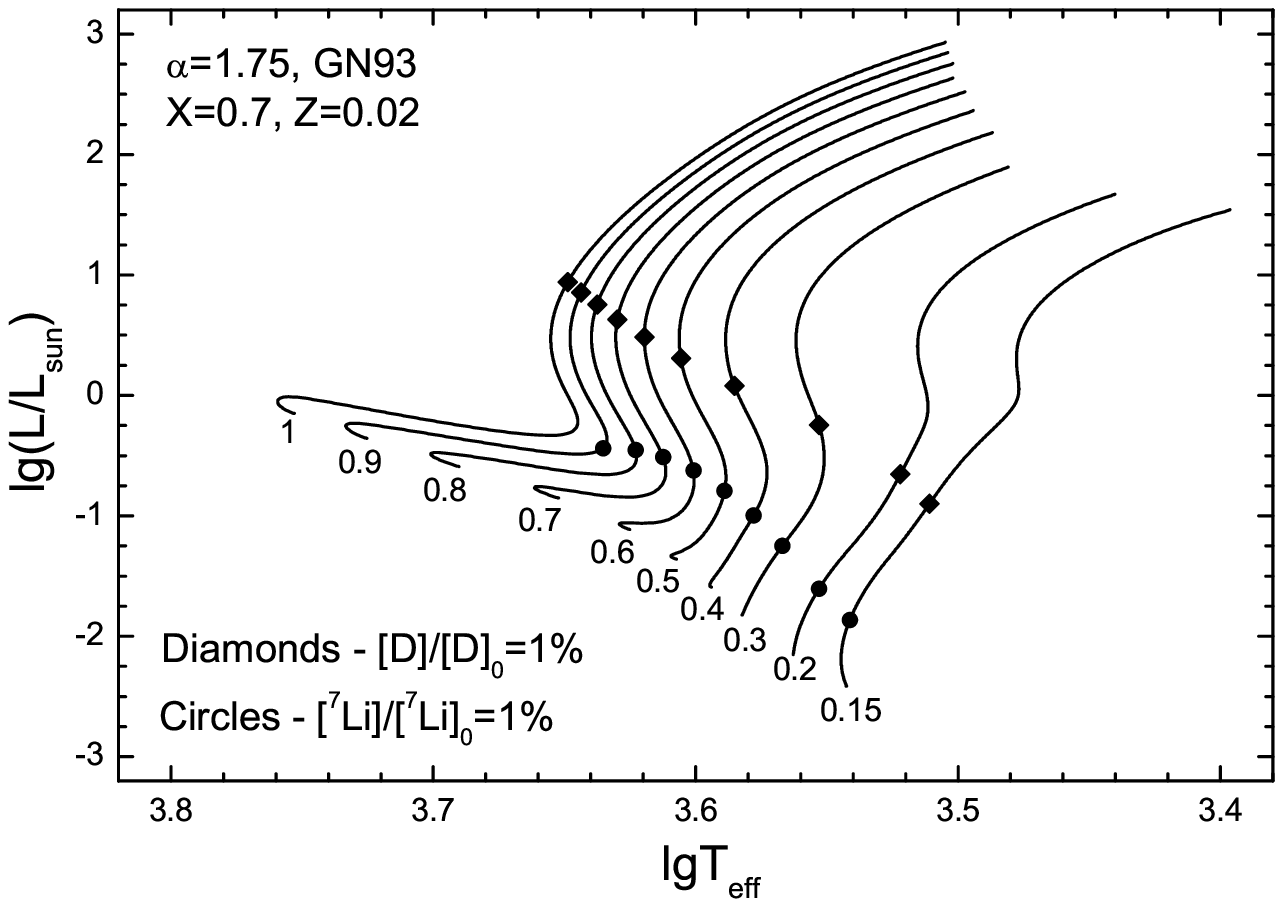}\vspace{1mm}
\caption{Evolutionary tracks of stars with $0.15\leq M/M_{\odot}\leq 1$ from PMS to ZAMS. X=0.7, Z=0.02 and MLT parameter $\alpha=1.75$ are used. Metal composition is same to GN93. Diamonds and circles denote the location of $[D]/[D]_0=1\%$ and $[^7Li]/[^7Li]_0=1\%$ at the stellar surface. The number below each track shows the mass of the star (in solar mass). } \label{depletion}
\end{figure*}

YNEV evolution code traces the variations of isotopes $D$ and $^7Li$, thus one can use it to study the depletion of light elements $D$ and $^7Li$ in the PMS stage. Figure \ref{depletion} shows the evolutionary tracks of stars with $0.15\leq /M_{\odot}\leq 1$ from PMS to ZAMS and denotes the depletions (to $1\%$ of initial abundance) of $D$ and $^7Li$ at the stellar surface. The stellar models are standard that only the fully mixing in the convective instable zone is taken into account and no extra mixing. Comparing with the MESA code \citep[see][Figure 15]{pax11}, the YNEV shows similar results, since both YNEV and MESA show same luminosity for the depletion points of $D$ / $^7Li$.

\subsection{Chemical composition in stellar interior: comparison with STAROX code}
\label{subsect:chem}

\begin{figure*}
\centering
\includegraphics[width=11cm]{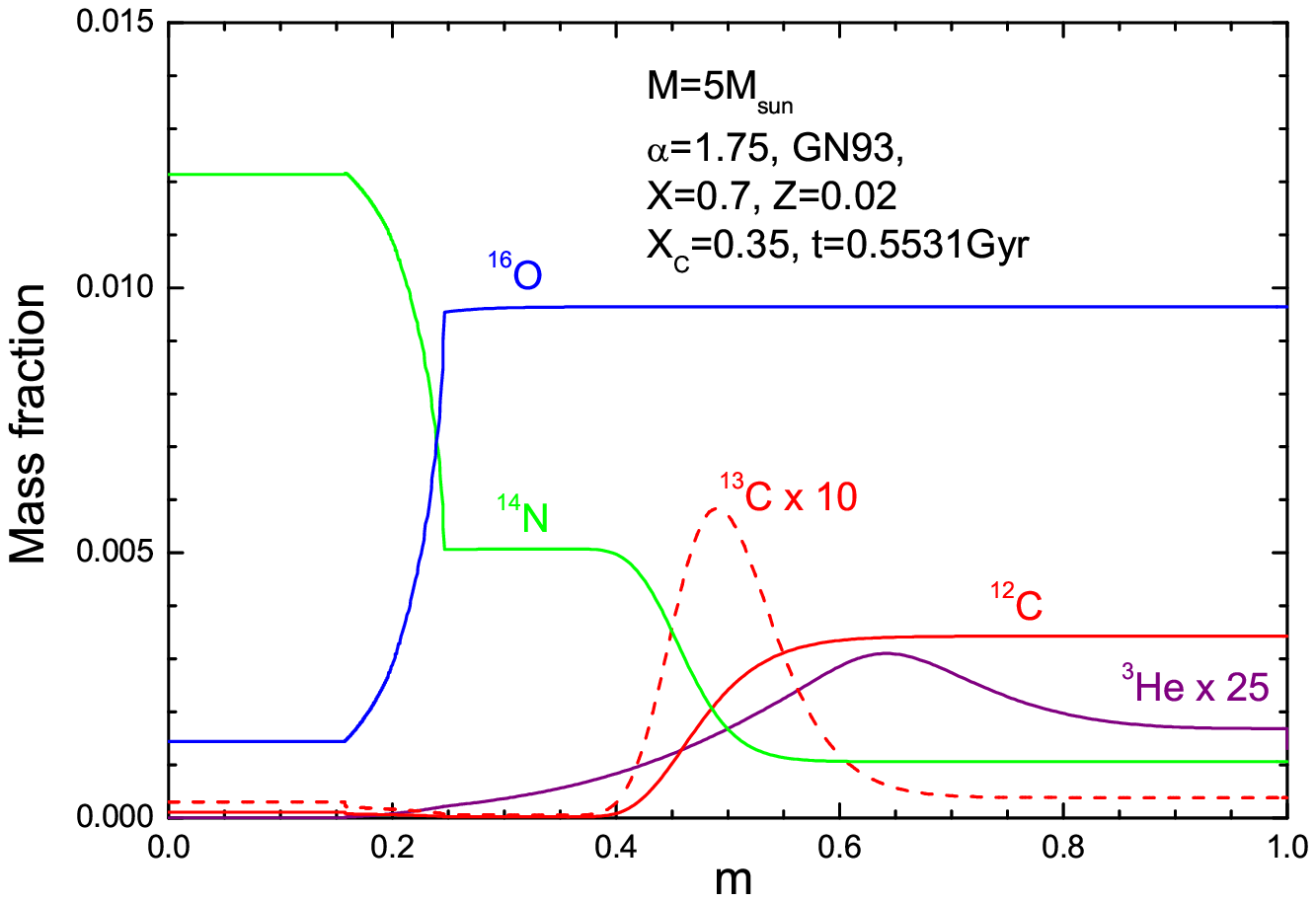}\vspace{1mm}
\caption{Chemical composition in the stellar interior of a stars with $M=5M_{\odot}$ and $X_C=0.35$. X=0.7, Z=0.02 and MLT parameter $\alpha=1.75$ are used. Metal composition is same to GN93. } \label{chem5m}
\end{figure*}

\begin{figure*}
\centering
\includegraphics[width=11cm]{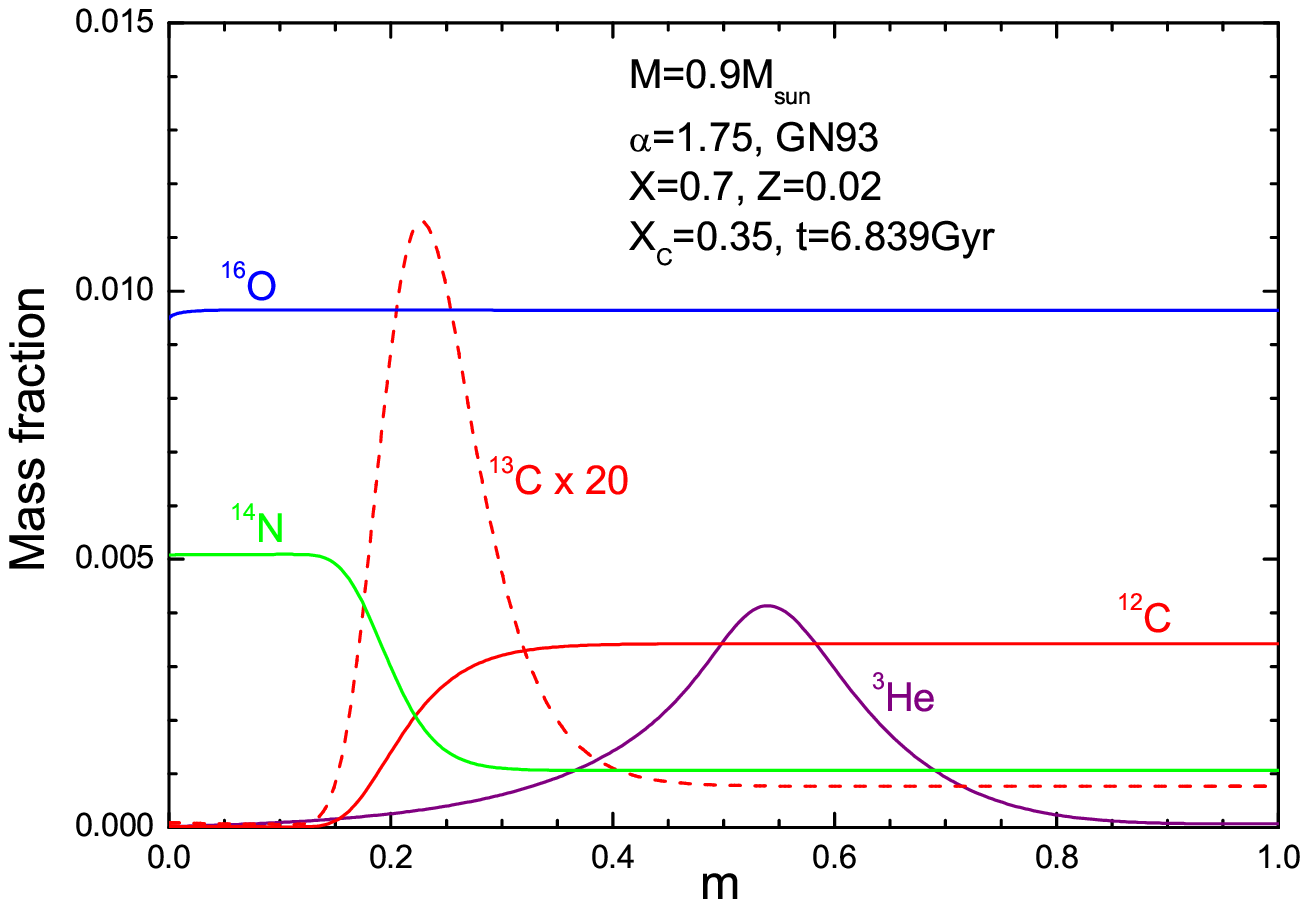}\vspace{1mm}
\caption{Similar to Fig.\ref{chem5m}, but for a $0.9M_{\odot}$ stars. } \label{chem09m}
\end{figure*}

Figures \ref{chem5m} \& \ref{chem09m} show the compositions in stellar interior of $0.9M_{\odot}$ and $5M_{\odot}$ stars with X=0.7, Z=0.02 at the stage of $X_C=0.35$. $\alpha=1.75$ and GN93 metal composition are used. The processes of $^{12}C, ^{16}O \rightarrow ^{14}N$ in the CNO-cycles are clear shown. $^{16}O \rightarrow ^{14}N$ can hardly occur in the $0.9M_{\odot}$ star since the temperature in the core is low. Those examples are also shown by the STAROX stellar evolution code \citep{rox08}. Comparing the YNEV results with STAROX's, there is no significant difference except for the age of the $0.9M_{\odot}$ star with $X_C=0.35$, which is $6.839Gyr$ in YNEV and $6.675Gyr$ in STAROX. The difference is though to be caused by the different initial abundance of the isotopes in metal and $^3He$.

\subsection{Evolutionary tracks in the the HR Diagram: for the nonlocal turbulent convection model}
\label{subsect:tracktcm}

\begin{figure*}
\centering
\includegraphics[width=11cm]{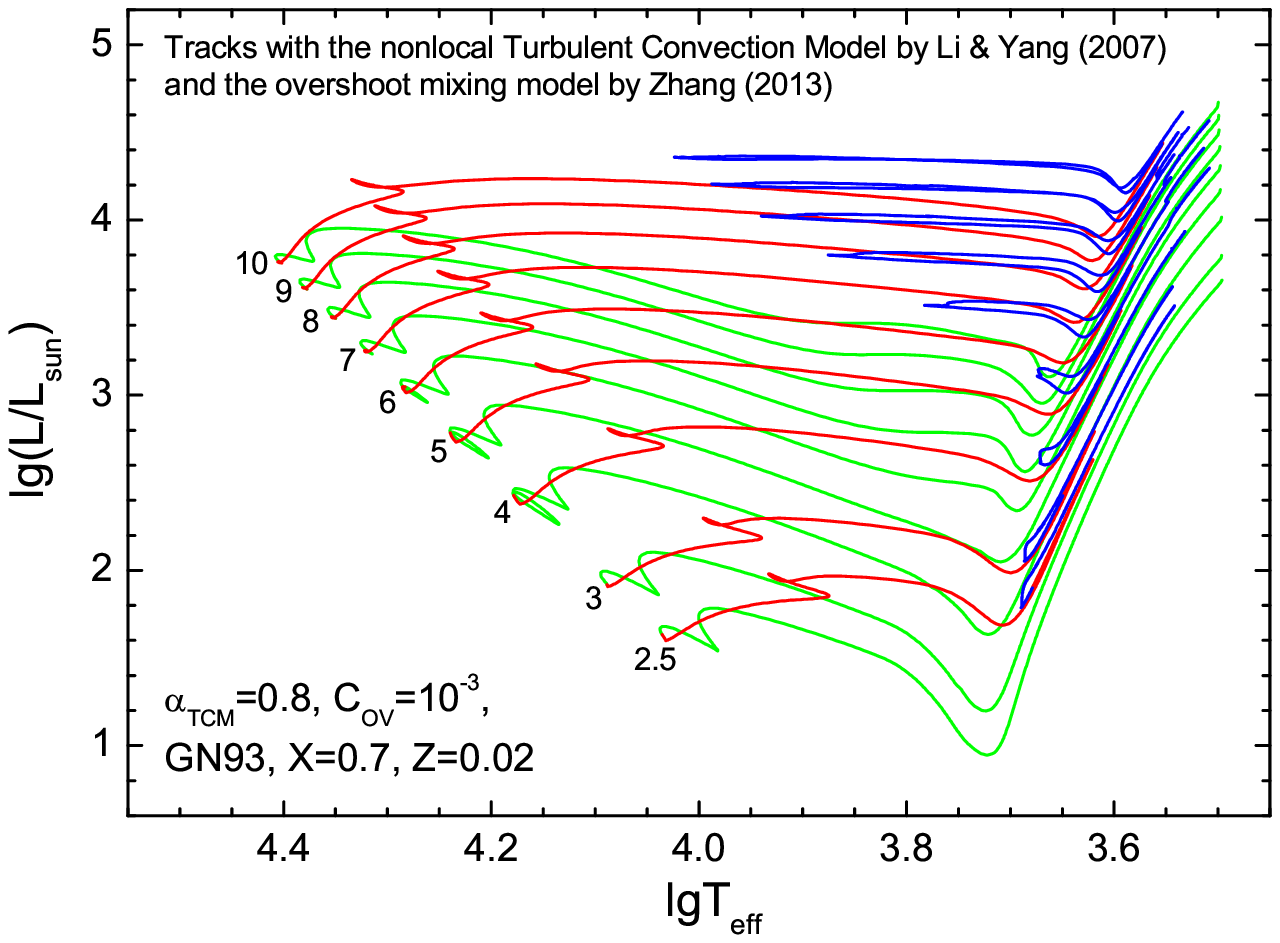}\vspace{1mm}
\caption{Evolutionary tracks of intermediate-mass stars with X=0.7, Z=0.02, the nonlocal turbulent convection model (convection parameter $\alpha_{TCM}=0.8$), and the updated overshoot model (with $C_{OV}=10^{-3}$) are used. Metal composition is same to GN93. In the tracks, the green, red and blue parts correspond to PMS, hydrogen burning and helium burning phases. The numbers near the ZAMSs indicate the stellar mass (in solar mass). } \label{ModelTCM1}
\end{figure*}

\begin{figure*}
\centering
\includegraphics[width=11cm]{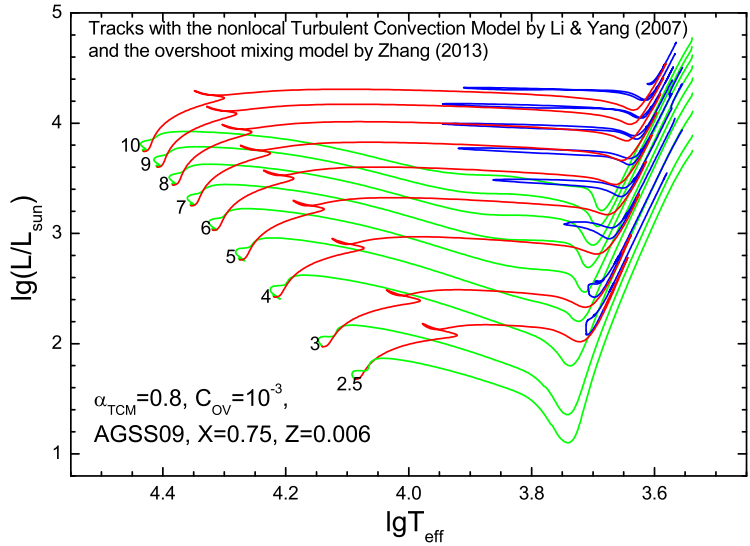}\vspace{1mm}
\caption{Similar to Fig.\ref{ModelTCM1}, but for X=0.76, Z=0.006, and the AGSS09 composition. } \label{ModelTCM2}
\end{figure*}

\begin{figure*}
\centering
\includegraphics[width=11cm]{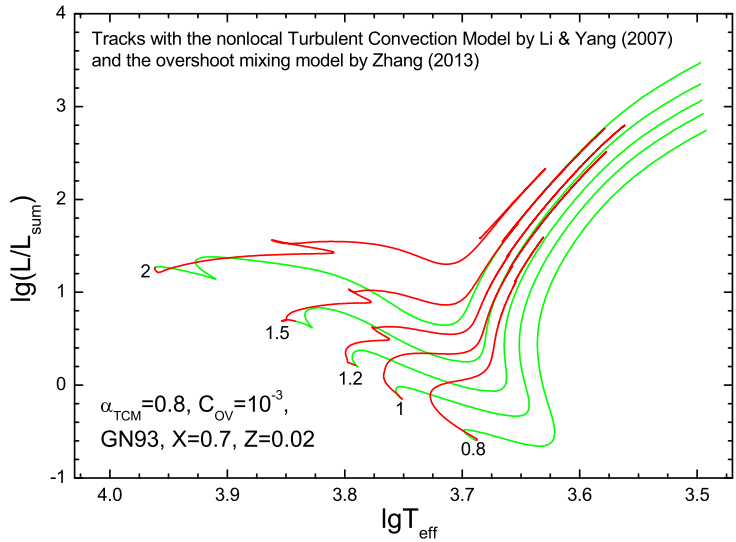}\vspace{1mm}
\caption{Similar to Fig.\ref{ModelTCM1}, but for low-mass stars. } \label{ModelTCM3}
\end{figure*}

The most important feature of the YNEV code is the ability to use the nonlocal turbulent convection theory in calculations of stellar structure and evolution. We show the evolutionary tracks of low- and intermediate-mass stars in Figures (\ref{ModelTCM1}-\ref{ModelTCM3}) for examples. The value of the turbulent kinetic dissipation parameter $\alpha_{TCM}$ is based on the solar calibration, and the value of the overshoot mixing parameter $C_{OV}$ is based on some observational restrictions \citep{zha13,mz14}. The stars evolves from the PMS with center temperature $T_C=10^5K$ to where the numerical scheme cannot find the solution satisfying both the TCM and the stellar structure equations. The localized TCM is used in the PMS with $lgT_C<6.8$, and the nonlocal TCM is used after that. It can be found that the scheme of implement of the nonlocal TCM works well in most cases of stellar evolution.

\subsection{Comparisons the nonlocal turbulent convection YNEV stellar model with Padova / Yale-Yonsei}
\label{subsect:comparison}

\begin{figure*}
\centering
\includegraphics[width=11cm]{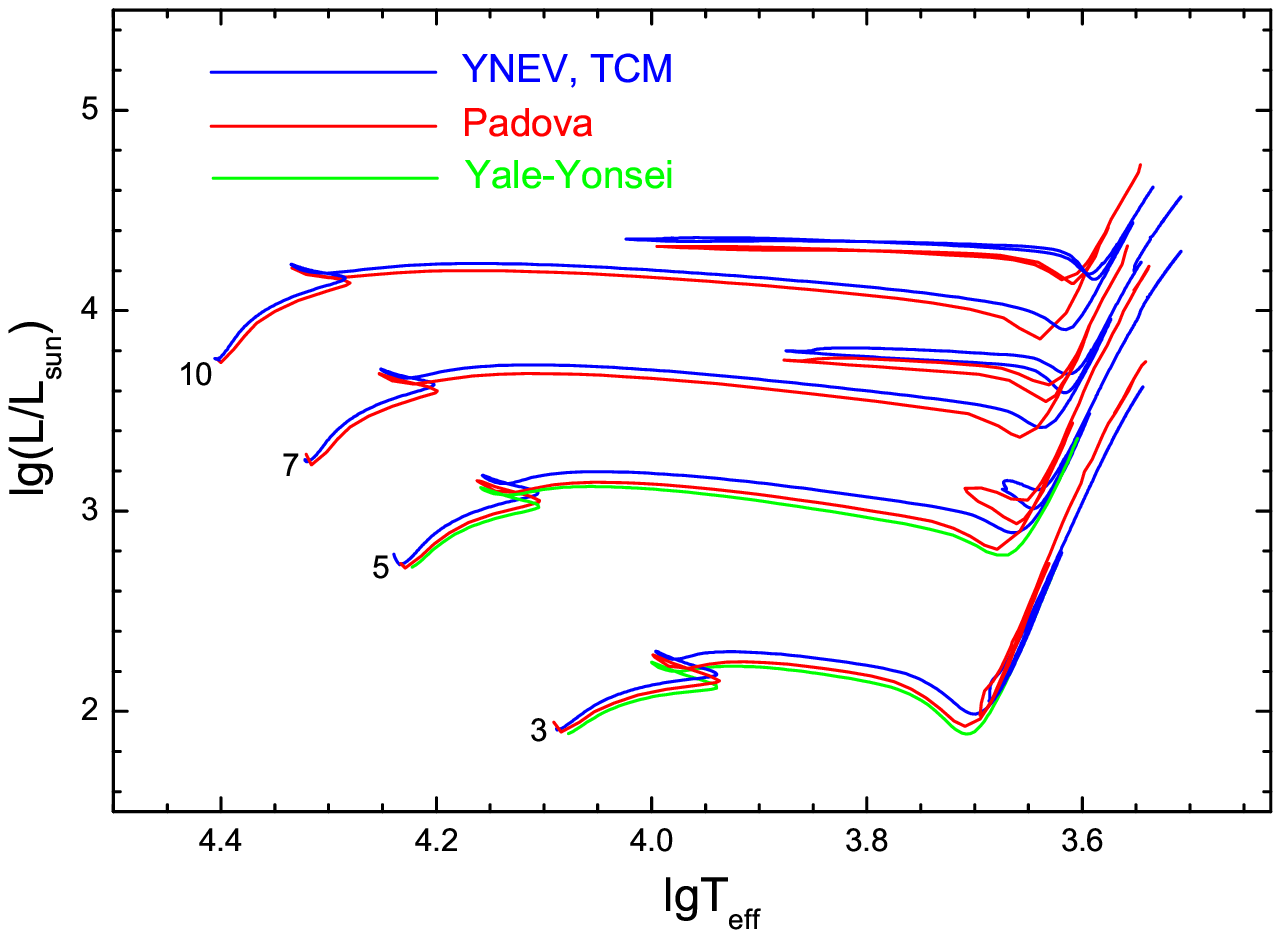}\vspace{1mm}
\caption{Padova, Yale-Yonsei, and YNEV inter-mediate mass star evolutionary tracks. The YNEV model is same as in Figure \ref{ModelTCM3}.}\label{comp1}
\end{figure*}

\begin{figure*}
\centering
\includegraphics[width=11cm]{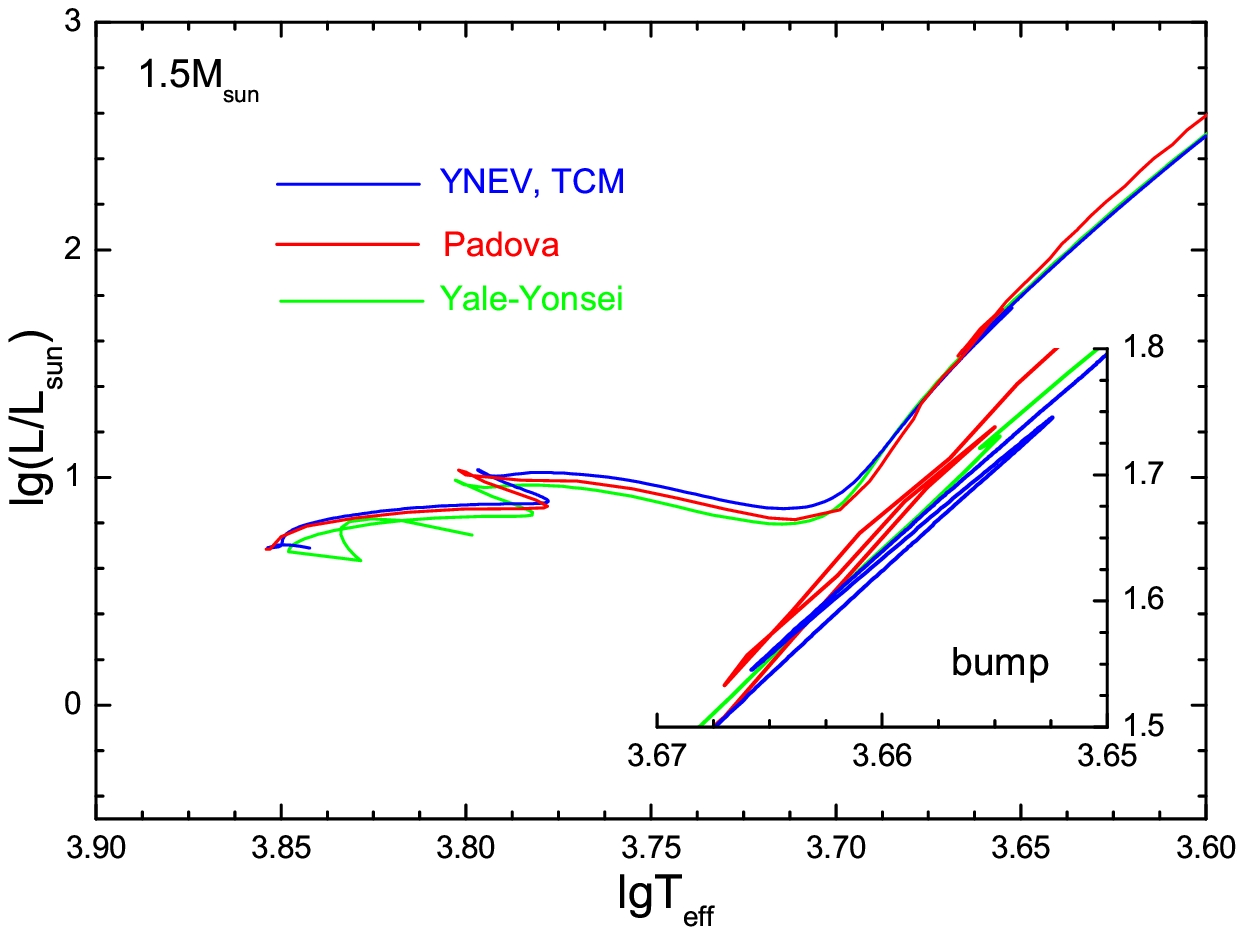}\vspace{1mm}
\caption{Similar to Figure \ref{comp1}, but for a $1.5M_{\odot}$ star. The bumps are zoomed. } \label{comp2}
\end{figure*}

Figure \ref{comp1} and \ref{comp2} show the evolutionary tracks of Padova \citep{PadovaTracks3,PadovaTracks1,PadovaTracks2}, Yale-Yonsei ($Y^2$) \citep{Y201,Y202,Y203,Y204}, and YNEV stellar models. The nonlocal TCM and the updated overshoot mixing model are used in the YNEV models, with $\alpha_{TCM}=0.8$ and $C_{OV}=10^{-3}$. The Padova stellar models are for X=0.708, Z=0.019, and the ballistic overshoot model \citep{bre81} with a stellar mass dependent overshoot parameter. The $Y^2$ stellar models are for X=0.71, Z=0.02, and fully mixed core overshoot region in $\alpha_{OV}H_P$ with the overshoot parameter $\alpha_{OV}$ being stellar mass dependent \citep{Y204}.

Figure \ref{comp1} shows the tracks for intermediate-mass stars. The main sequence band widthes of three models indicate that the core overshoot mixing strength in three code are similar. The extensions of the blue loops in YNEV and Padova are similar. Comparing with the Padova tracks, YNEV tracks are at lower temperature in the RGB phase and the differences increases as the stellar mass increasing. It should be noticed that the low-temperature opacity table in the YNEV is different from that in the Padova, i.e., \citeauthor{F05}'s \citeyearpar{F05} tables are adopted in YNEV and \citeauthor{af94}'s \citeyearpar{af94} tables are adopted in Padova. And the efficiency of the turbulent heat transport in the super-adiabatic convection zone shows difference between the TCM and the MLT.
Figure \ref{comp2} shows the case of the $1.5M_{\odot}$ star. The main difference is that YNEV and Padova models show large bump in the RGB phase, but the $Y^2$ model shows small bump. We think that it is caused by the absence of the overshoot mixing below the convective envelope in the $Y^2$ model.

Table 2 shows the age of critical points in stellar evolutionary tracks for Padova, $Y^2$, and YNEV (with nonlocal TCM) stellar models. It is found that the age of critical points in YNEV stellar models with nonlocal TCM and the updated overshoot are the same as Padova and $Y^2$, except the MS age of the low-mass star. The MS age for the YNEV stellar models is a little larger than that of Padova and $Y^2$. Those results shows that the core overshoot mixing strength in three code are similar and, for low-mass star, YNEV shows a little stronger overshoot mixing. It is noticed that the overshoot parameters in Padova and $Y^2$ are stellar mass dependent but the overshoot parameter in the YNEV is constant.

\begin{table*}
\centering
\caption{The age (in $Gyr$) of critical points in stellar evolutionary. (Y2), (P) and (YN) means for Yale-Yonsei, Padova and YNEV stellar models, respectively. }
\begin{tabular}{lccccc}
\hline\hline\noalign{\smallskip}
critical point  & 1.5M & 3M & 5M & 7M & 10M \\
\hline\noalign{\smallskip}
$H$ burned out in the center   & 2.895 (Y2)  & 0.3921 (Y2)  & 0.1040 (Y2) & - (Y2)       & - (Y2)          \\
                               & 2.755 (P)   & 0.3793 (P)   & 0.1031 (P)  & 0.04784 (P)  & 0.02379 (P)    \\
                               & 3.175 (YN)  & 0.3855 (YN)  & 0.1023 (YN) & 0.04735 (YN) & 0.02350 (YN)   \\
$He$ ignition in the center    & - (Y2)      & - (Y2)       & - (Y2)      & - (Y2)       & - (Y2)          \\
                               & - (P)       & 0.3834 (P)   & 0.1037 (P)  & 0.04803 (P)  & 0.02384 (P)    \\
                               & - (YN)      & 0.3905 (YN)  & 0.1031 (YN) & 0.04759 (YN) & 0.02357 (YN)   \\
$He$ burned out in the center  & - (Y2)      & - (Y2)       & - (Y2)      & - (Y2)       & - (Y2)          \\
                               & - (P)       & 0.4763 (P)   & 0.1174 (P)  & 0.05277 (P)  & 0.02585 (P)    \\
                               & - (YN)      & 0.4775 (YN)  & 0.1167 (YN) & 0.05293 (YN) & 0.02608 (YN)   \\
\hline
\end{tabular}
\end{table*}

\subsection{Stellar oscillations}
\label{subsect:osc}

\begin{figure*}
\centering
\includegraphics[width=11cm]{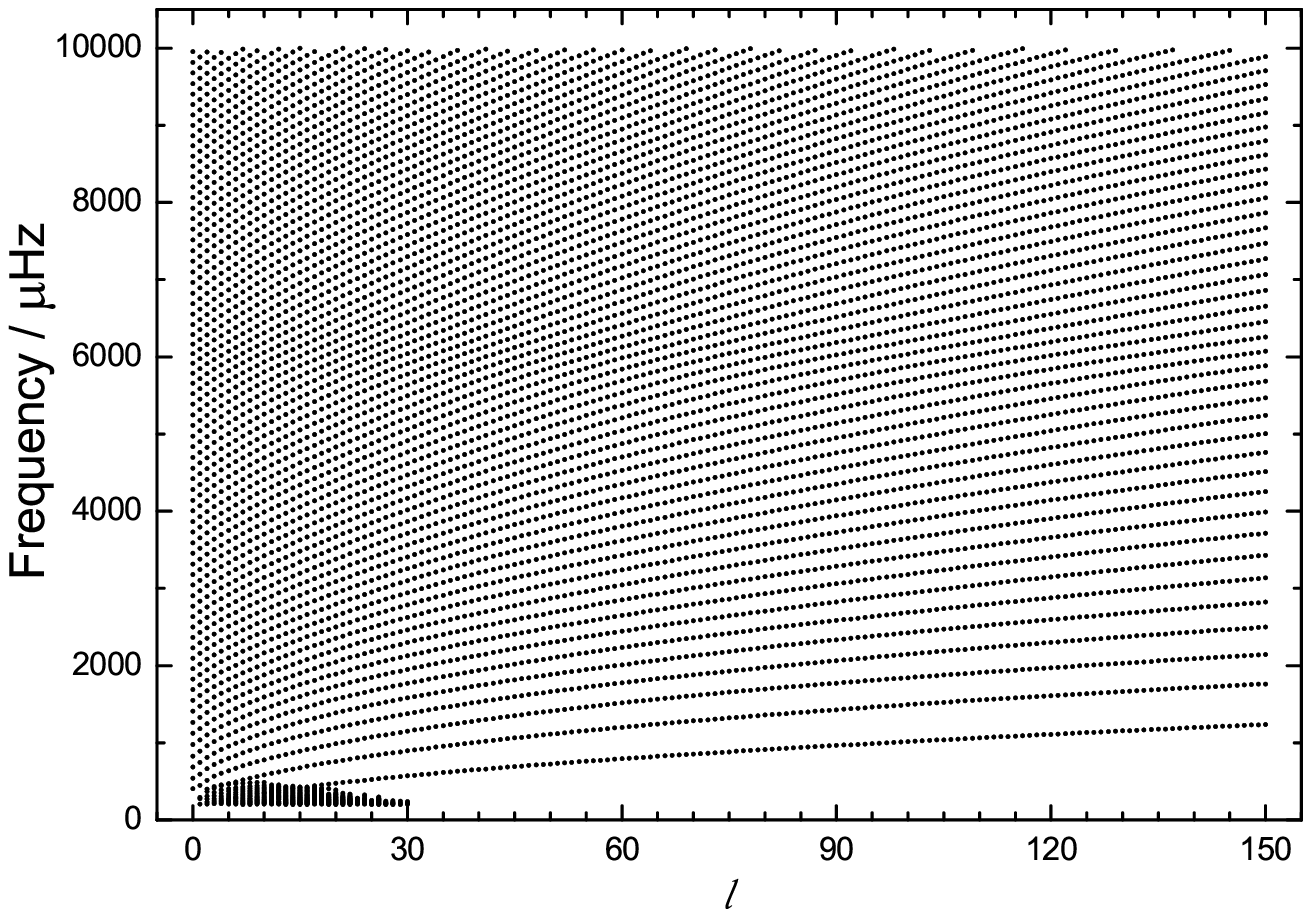}\vspace{1mm}
\caption{Eigenfrequencies in the range of $200\leq f/\mu Hz \leq 10000$ of the solar model Z93 for $0\leq l\leq150$.} \label{solarLF}
\end{figure*}

\begin{figure*}
\centering
\includegraphics[width=11cm]{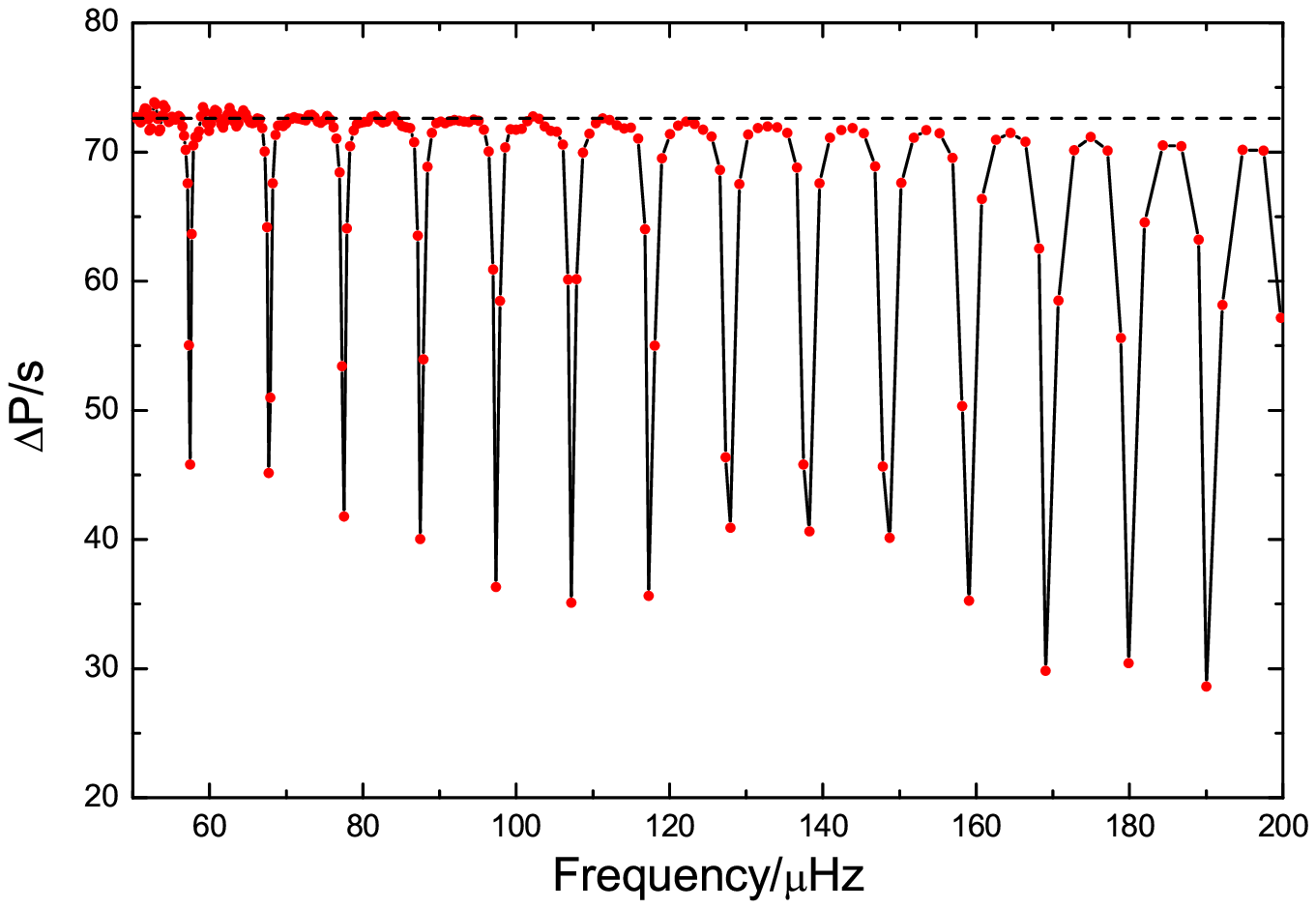}\vspace{1mm}
\caption{Relation between period spacing and frequency of the $1.5M_{\odot}$ RGB star with $R=6.38R_{\odot}$ and $L=19.3L_{\odot}$. The dashed line shows the asymptotical value $\Delta P=72.61s$. } \label{RGBosc}
\end{figure*}

The linear adiabatic oscillation part of YNEV code is designed to scan the eigenfrequencies and to solve the eigenfunctions of stellar adiabatic oscillations.
We show here two samples of the applications of YNEV oscillation code: oscillations of the solar model and mixed modes of a RGB low-mass star. Figure \ref{solarLF} shows the eigenfrequencies of the solar model Z93 calculated by using the YNEV oscillation code. The eigenfrequencies are in the range of $200\leq f/\mu Hz \leq 10000$ for $0\leq l\leq150$. The dense part in low-frequencies for low $l$ are g-modes and others are p-modes.
Figure \ref{RGBosc} shows the relation between period spacing $\Delta P$ and frequency $f$ of the $1.5M_{\odot}$ RGB star with the radius $R=6.38R_{\odot}$ and the luminosity $L=19.3L_{\odot}$. A similar sample was studied by \cite{bed11} using the ASTEC evolution code \citep{chr08a} and the ADIPLS oscillation code \citep{chr08b}.

\section{Developments in the future}
\label{sect:devlp}

There are many ways the code could be improved. In order to calculate massive stars, more nuclear burning reactions should be included, the possible H-semiconvection zone and convective burning shell(s) out side the convective core should be treated properly. At present, the stellar structure and the composition are solved individually. A better way is to solve them together. The two point 2nd order discretisation
could be upgraded to higher order scheme. An important improvement we planed is to apply parallel numerical calculations, which can significantly boosts the calculation speed. To my knowledge, the YNEV is the first stellar evolution code which could use a nonlocal turbulent convection model in the most cases of stellar evolution. As mentioned above, there are still some numerical problem of the implement of the nonlocal TCM. Correcting those numerical problem is a priority. It is no doubt that reader can think of other ways in which the
code could be improved.

\normalem
\begin{acknowledgements}

QSZ thanks a referee for constructive comments which
improved the original manuscript. This work is co-sponsored by the National Natural Science
Foundation of China (NSFC) through grant No. 11303087, the West Light Foundation of the Chinese Academy
of Sciences, the Science Foundation of Yunnan Observatory No. Y1ZX011007 \& Y3CZ051005 and the Chinese Academy
of Sciences under grant No. KJCX2-YW-T24. Fruitful
discussions with Yan Li and J{\o}rgen Christensen-Dalsgaard are highly appreciated.

\end{acknowledgements}

%\bibliographystyle{raa}
%\bibliography{bibtex}

\clearpage

\begin{thebibliography}{99}

  \bibitem[Alexander \& Ferguson (1994)]{af94} Alexander, D. R., \& Ferguson, J. W. 1994, \apj, 437, 879

  \bibitem[Angulo (1999)]{ang99} Angulo, C., et al. 1999, Nucl. Phys. A, 656, 3

  \bibitem[Aparicio (1998)]{apa98} Aparicio, J. M., \& Arnett, D. 1998, \apjs, 117, 627

  \bibitem[Asplund et al. (2009)] {AGSS09} Asplund, M., Grevesse, N., Sauval, A. J.,\& Scott, P. 2009, ARA\&A, 47, 481

  \bibitem[Basu et al. (2009)] {ba09} Basu, S., Chaplin, W. J., Elsworth, Y., New, R., \& Serenelli, A. M. 2009, \apj, 669, 1403

  \bibitem[Bedding et al. (2011)] {bed11} Bedding, T. R., Mosser, B., Huber, D., Montalb\'{a}n, J., Beck, P., \& Christensen-Dalsgaard, J., et al. 2011, Nature, 471, 608

  \bibitem[Bressan et al. (1981)] {bre81} Bressan, A., Bertelli, G., \& Chiosi, C., 1981, \aap, 102, 25

  \bibitem[Bono et al. (2000)] {FRANECBlueloop} Bono, G., Caputo, F., Cassisi, S., et al. 2000, \apj, 543, 955

  \bibitem[B\"{o}m-Vitense (1958)] {MLT58} B\"{o}m-Vitense, E. 1958, Z. Astrophys., 46, 108

  \bibitem[Canuto (1997)] {can97} Canuto, V. M. 1997, \apj, 482, 827

  \bibitem[Canuto \& Dubovikov (1998)] {can98} Canuto, V. M., \& Dubovikov, M. 1998, \apj, 493, 834

  \bibitem[Canuto (2011)] {can11} Canuto, V. M. 2011, \aap, 528, 76

  \bibitem[Cassisi (2007)] {cas07} Cassisi, S., Potekhin, A. Y., Pietrinferni, A., Catelan, M., \& Salaris, M. 2007, \apj, 661, 1094

  \bibitem[Castellani et al. (1985)] {cas85}  Castellani, V., Chieffi, A., Pulone, L., \& Tornamb\`{e}, A. 1985, \apj, 296, 204

  \bibitem[Caughlan \& Fowler (1988)]{cf88} Caughlan, G. R., \& Fowler, W. A. 1988, At. Data Nucl. Data Tables, 40, 283

  \bibitem[Christensen-Dalsgaard (2008a)] {chr08a}  Christensen-Dalsgaard, J., 2008a, \apss, 316, 13

  \bibitem[Christensen-Dalsgaard (2008b)] {chr08b}  Christensen-Dalsgaard, J., 2008b, \apss, 316, 113

  \bibitem[Christensen-Dalsgaard et al. (2011)] {chr11}  Christensen-Dalsgaard, J., Monteiro, M. J. P. F. G., Rempel, M., \& Thompson, M. J. 2011, \mnras, 414, 1158

  \bibitem[de Jager et al. (1998)] {dej98}  de Jager, C., Nieuwenhuijzen, H., \& van der Hucht, K. A., 1988, A\&AS, 72, 259

  \bibitem[Demarque et al. (2004)] {Y204}  Demarque, P., Woo, J. H., Kim, Y. C., \& Yi, S. K. 2004, \apjs, 155, 667

  \bibitem[Deng \& Xiong (2006)] {den06} Deng, L., Xiong, D. R., \& Chan, K. L. 2006, \apj, 643, 426

  \bibitem[DeWitt et al. (1973)]{dgc73} DeWitt, H. E., Graboske, H. C., \& Cooper, M. S. 1973, \apj, 181, 439

  \bibitem[Ferguson et al. (2005)] {F05} Ferguson, J. W., Alexander, D. R., Allard, F., Barman, T., Bodnarik, J. G., Hauschildt, P. H., Heffner-Wong, A., \& Tamanai, A. 2005, \apj, 623, 585

  \bibitem[Grevesse \& Noels (1993)] {GN93} Grevesse, N., \& Noels, A., 1993, in Prantzos N., Vangioni-Flam E., Casse M.,
  eds, Origin and Evolution of the Elements. Cambridge Univ. Press,
  Cambridge, p. 15

  \bibitem[Grevesse \& Sauval (1998)] {GS98} Grevesse, N., \& Sauval, A. J., 1998, Space Sci. Rev., 85, 161

  \bibitem[Iben (1975)] {iben75} Iben, I. 1975, \apj, 196, 525

  \bibitem[Iglesias \& Rogers (1996)] {ig96} Iglesias, C. A., \& Rogers, F. J. 1996, \apj, 464, 943

  \bibitem[Itoh et al. (1996)]{ito96} Itoh, N., Hayashi, H., Nishikawa, A., \& Kohyama, Y. 1996, \apjs, 102, 411

  \bibitem[Kim et al. (2002)] {Y202}  Kim, Y. C., Demarque, P., Yi, S. K., \& Alexander, D. R. 2002, \apjs, 143, 499

  \bibitem[Krishna Swamy (1966)] {ks66} Krishna Swamy, K. S. 1966, \apj, 145, 174

  \bibitem[Li \& Yang (2007)] {li07} Li, Y.,\& Yang, J. Y. 2007, \mnras, 375, 388

  \bibitem[Li (2010)] {li10} Li, Y. 2010, Stellar oscillation lectures

  \bibitem[Li (2012)] {li12} Li, Y. 2012, \apj, 756, 37

  \bibitem[Meng \& Zhang (2014)] {mz14} Meng, Y., \& Zhang, Q. S. 2014, \apj, acepted

  \bibitem[Paczynski (1969)] {pacz69} Paczynski, B. 1969, AcA, 19, 1

  \bibitem[Paxton et al. (2011)] {pax11} Paxton, B., Bildsten, L., Dotter, A., Herwig, F., Lesaffre, P., \& Timmes, F. 2011, \apjs, 192, 3

  \bibitem[Reimers (1975)] {rei75} Reimers, D., 1975, MSRSL, 8, 369

  \bibitem[Rogers \& Nayfonov (2002)]{RN02} Rogers, F. J., \& Nayfonov, A. 2002, \apj, 576, 1064

  \bibitem[Roxburgh (2008)] {rox08} Roxburgh I. W. 2008, \apss, 316, 75

  \bibitem[Salasnich et al. (2000)] {PadovaTracks1} Salasnich, B., Girardi, L., Weiss, A., \& Chiosi, C., 2000, \aap, 361, 1023

  \bibitem[Girardi et al. (2000)] {PadovaTracks2} Girardi, L., Bressan, A., Bertelli, G., \& Chiosi, C., 2000, A\&AS, 141, 371

  \bibitem[Bressan et al. (1993)] {PadovaTracks3} Bressan A., Fagotto F., Bertelli G., \& Chiosi C., 1993, A\&AS, 100,647

  \bibitem[Salpeter (1954)]{sal54} Salpeter, E.E. 1954, Aust. J. Phys. 7, 373

  \bibitem[Seaton (2005)] {OP05} Seaton, M. J. 2005, \mnras, 362, 1

  \bibitem[Timmes \& Arnett (1999)]{ta99} Timmes, F. X., \& Arnett, D. 1999, \apjs, 125, 277

  \bibitem[Thoul et al. (1994)] {tbl94} Thoul, A. A., Bahcall, J. N., \& Loeb, A. 1994, \apj, 421, 828

  \bibitem[Waldron (1984)] {wal84}  Waldron, W.L., 1984, in A. B. Underhill and A. G. Michalitsianos (Eds.)
    The Origin of Non-Radiative Heating/Momentum in Hot Stars, NASA Washington D.C., p. 2358

  \bibitem[Xiong (1981)] {xio81} Xiong, D. R. 1981, Sci. Sinica, 24, 1406

  \bibitem[Xiong et al. (1997)] {xio97} Xiong, D. R., Cheng, Q. L., Deng, L. 1997, \apjs, 108, 529

  \bibitem[Yakovlev \& Urpin (1980)] {yak80} Yakovlev, D. G, \& Urpin, V. A. 1980, SvA, 24, 303

  \bibitem[Yi et al. (2001)] {Y201}  Yi, S. K., Demarque, P., Kim, Y. C., et al. 2001, \apjs, 136, 417

  \bibitem[Yi et al. (2003)] {Y203}  Yi, S. K., Kim, Y. C., \& Demarque, P. 2003, \apjs, 144, 259

  \bibitem[Zhang \& Li (2012)]{zhl12} Zhang, Q. S., \& Li, Y. 2012, \apj, 746, 50

  \bibitem[Zhang (2012)]{zha12} Zhang, Q. S. 2012, \apj, 761, 153

  \bibitem[Zhang (2013)]{zha13} Zhang, Q. S. 2013, \apjs, 205, 18

  \bibitem[Zhang (2014)]{zha14} Zhang, Q. S. 2014, ApJL, 787, 28

\end{thebibliography}

%\appendix

\label{lastpage}

\end{document}